\documentclass[twocolumn]{IEEEtran}
\usepackage{hyperref}
\usepackage{lettrine}
\usepackage{epsfig}
\usepackage{footnote}
\usepackage{cite}
\usepackage{float}
\usepackage{amssymb}
\usepackage{amsthm}
\usepackage{amsmath,mathabx}
\usepackage{color}
\usepackage{amsfonts}
\usepackage{epstopdf}
\usepackage{tabularx,tabulary}
\usepackage[font=footnotesize]{caption}
\usepackage{graphicx}
\usepackage{caption}
\usepackage{subcaption}
\usepackage{wrapfig}
\usepackage{dblfloatfix}    % To enable figures at the bottom of page
\usepackage{bbm}
\usepackage[dvipsnames]{xcolor}
\usepackage{algorithm,algpseudocode}

\newcommand*\rfrac[2]{{}^{#1}\!/_{#2}}
\newcommand{\vect}[1]{\boldsymbol{#1}}

\DeclareMathOperator{\argmax}{argmax}

\newtheorem{mydef}{Definition}

\definecolor{kirmizi}{RGB}{204,0,0}

\begin{document}
\title{Design and Provision of Traffic Grooming for Optical Wireless Data Center Networks}
\author{Abdulkadir~Celik,~\IEEEmembership{Member,~IEEE,} Amer~AlGhadhban,~\IEEEmembership{Student Member,~IEEE,} \\
Basem Shihada,~\IEEEmembership{Senior Member,~IEEE,} and Mohamed-Slim~Alouini,~\IEEEmembership{Fellow,~IEEE}.

\thanks{The authors are with Computer, Electrical, and Mathematical Sciences and Engineering (CEMSE) Division at King Abdullah University of Science and Technology (KAUST), Thuwal, 23955-6900, KSA. A part of this paper was presented in IEEE WCNC 2018 in Barcelona, Spain \cite{Celik18design}. Corresponding author: Abdulkadir Celik (abdulkadir.celik@kaust.edu.sa)}
}
\markboth{IEEE Transactions on Communications}{Celik \MakeLowercase{\textit{et al.}}: Design and Provision of Traffic Grooming for Optical Wireless Data Center Networks}

\maketitle
\begin{abstract}
Traditional wired data center networks (DCNs) suffer from cabling complexity, lack flexibility, and are limited by the speed of digital switches. In this paper, we alternatively develop a top-down traffic grooming (TG) approach to the design and provisioning of mission-critical optical wireless DCNs. While switches are modeled as hybrid optoelectronic cross-connects, links are modeled as wavelength division multiplexing (WDM) capable free-space optic (FSO) channels. Using the standard TG terminology, we formulate the optimal mixed-integer TG problem considering the virtual topology, flow conversation, connection topology, non-bifurcation, and capacity constraints. Thereafter, we develop a fast yet efficient sub-optimal solution which grooms mice flows (MFs) and mission-critical flows (CFs) and forward on predetermined rack-to-rack (R2R) lightpaths. On the other hand, elephant flows (EFs) are forwarded over dedicated server-to-server (S2S) express lightpaths whose routes and capacity are dynamically determined based on the availability of wavelength and capacity. To prioritize the CFs, we consider low and high priority queues and analyze the delay characteristics such as waiting times, maximum hop counts, and blocking probability. As a result of grooming the sub-wavelength traffic and adjusting the wavelength capacities, numerical results show that the proposed solutions can achieve significant performance enhancement by utilizing the bandwidth more efficiently, completing the flows faster than delay sensitivity requirements, and avoiding the traffic congestion by treating EFs and MFs separately.
\end{abstract}
\IEEEpeerreviewmaketitle

\section{Introduction}
\label{sec:intro}

\lettrine{D}{ata centers} (DCs) are becoming an intrinsic part of the computing infrastructures for emerging technologies which requires storage and processing of massive amounts of data. While some enterprises and governmental institutions build and sustain their own DCs, some others meet these demands by renting or purchasing a portion of large-scale DC networks (DCNs) offered by gigantic technology firms \cite{Celik18design}. DCN applications include but not limited to cloud services, big data, artificial intelligence, content delivery, neural networks, and cellular infrastructure; all of which necessitates sophisticated and interconnected storage and computing resources. For instance, cloud radio access is provisioned to be an enabler of the fifth generation (5G) networks by centralized processing of a massive number of devices \cite{peng2014heterogeneous}. Therefore, it is also necessary to be able to handle highly diversified types of traffic as a regular phone call and blue light box call cannot be treated in the same priority.

Scalability of DCNs is required to accommodate a large number of servers with adequate speeds and bandwidths. However, today's DCNs interconnects network equipment via unshielded twisted pair cables or fiber-optic wires, which has the following disadvantages; cabling cost and complexity, lack of flexibility,  and underutilization of the available bandwidth \cite{celikWDCNmagazine}. Fortunately, wired DCNs can be augmented with wireless technologies such as multi-gigabit mmWave \cite{Halperin2011augmenting} or multi-terabit free-space optic (FSO) \cite{Hamedazimi2014firefly}. While mmWave technology offers some degree of penetration, it suffers from interference and short ranges due to high attenuation.  Thanks to the line of sight (LoS) links among the receivers, FSO can alternatively provide an interference-free communication and naturally improve the physical layer security. However, this necessitates replacing the traditional grid-based rack arrangement of wired DCNs with a physical topology design that facilitates the LoS links. It is shown that an outdoor WDM-FSO link can achieve 1.28 Tbps (32x40 Gbps) capacity on 32 wavelengths over 212 meters distance \cite{Ciaramella2009128}. 

Unlike the outdoor FSO links which suffer from hostile channels impediments (e.g., scintillation, pointing error, and atmospheric turbulence, etc.), the quality of indoor FSO links within DCNs is primarily determined by the link distance as they operate in an acclimatized environment. Therefore, indoor FSO links can offer higher bandwidth at longer distances and the bandwidth can be further enhanced with wavelength division multiplexing (WDM) methods \cite{CelikSPIE}. In \cite{Hamza2014free}, an FSO based DCN design is presented based on fixed, non-mechanical, FSO links by means of fully connected rows/columns of racks. A more generalized OWCells concept is proposed by Hamza et. al. in \cite{Hamza2017} where fixed LoS optical wireless communication (OWC) links are employed to interconnect racks arranged in regular polygonal topologies. Readers can also refer to \cite{7393451} for a survey of recent advances in physical topology realizations in wireless DCNs.

Regardless of the potentially achievable multi-terabit link capacities by combining WDM and FSO, the DCN bottleneck is still determined by data processing capability of power-hungry state-of-art switches which can handle 1-10 Gbps rate at each port. Alternatively, power consumption and processing limitations can be mitigated by optical or hybrid optoelectronic switches. As the flow bandwidth requests can be much lower than the capacity offered by WDM channels, \textit{traffic grooming} (TG) arises as a necessary operation which refers to the aggregation of subwavelength flows onto high-speed lightpaths subject to equipment costs and capacity \cite{Huang2007dynamic}. Considering the enormous number of flows generated by servers across the DCN, TG is also vital to alleviate the network management complexity by combining the same class of flows and provision their common QoS demands altogether. However, TG poses daunting challenging as its subproblems have been shown to be NP-hard \cite{Zhur2002mesh}. Accounting for dynamically changing traffic conditions of DCNs, these entail developing fast yet efficient suboptimal TG policy designs. Accordingly, our goal in this paper is design and provision of WDM-FSO based WDCNs from a TG approach which is aware of various quality of service (QoS) domains including flow size, delay sensitivity, and priority.

\subsection{Related Work}
\label{sec:related}
Recent efforts on WDCNs can be exemplified as follows:  The idea of using 60 GHz technology in DCNs is first shared in \cite{ranachandran200860ghz} where Ramachandran et. al. identified the problems of wired DCNs and discussed the potentials and challenges of using 60 GHz band. In \cite{Halperin2011augmenting}, results of 60 GHz link measurement and simulation showed that directional links are essential for link stability, interference mitigation, and higher throughput. In \cite{Cui2011channel, Cui2011wireless, Cui2013dynamic}, Cui et. al. addressed the wireless channel allocation and scheduling to overcome congestion and interference of hybrid 60 GHz DCNs. On the other hand, interference mitigation is obtained by steered-beam mmWave links in wireless packet-switching networking DCNs \cite{Katayama2011steered}.

Riza et. al. proposed mechanically steerable FSO links in \cite{Riza2012power} where a transceiver is mounted to a pedestal platform which allows for vertical and rotational motion to establish links among the racks. In \cite{Hamedazimi2014firefly}, inter-rack communications are realized via FSO links reflected on a ceiling mirror similar to Flyways in \cite{Halperin2011augmenting}. Likewise, Bao et. al. presented FlyCast FSO DCN in \cite{Bao2015flycast} where they employ splitting state of the switchable mirrors to enable multicast without the need for a switch. In \cite{Arnon2015next}, Arnon developed a physical DCN topology to establish intra-rack and inter-rack communications. We refer interested readers to \cite{7393451} where Hamza et. al. surveyed the recent efforts on WDCNs under the classification of 60 GHz and FSO communications and compare virtues and drawbacks of potential wireless technologies for DCNs. 

Since power consumption and speed limit are two major drawbacks of packet-switched architectures, power reduction and throughput enhancement have been obtained by optical switches which may operate in time, space and wavelength domains. For example, all-optical switches can reach Petabit-scales by employing an arrayed waveguide grating router (AWGR) with hundreds of ports and hundreds of wavelength each with tens of Gbps data rate \cite{Yoo2013Elastic}. Alternatively, circuit switching based hybrid architectures are proposed to use semiconductor optical amplifiers or micro-electro-mechanical switches in \cite{Farrington:2010:Helios} and \cite{Wang:2010:cThrough}, respectively. Also, a mixed switching architecture is proposed in \cite{Fiorani2014energy} which combines wavelength routing, circuit switching, and broadcast-and-select mechanisms.
 
In the past decades, TG was extensively studied for synchronous optical networks for a variety of topologies (please refer to \cite{Huang2007dynamic} and references therein). However, to the best of our knowledge, its potentials and prospects are not studied in the realm of DCNs except in \cite{Sankaran2014scheduling, Sankaran2016optical} where traffic is simply groomed into three classes of wavelengths which are confined to broadcasting within racks and higher layer switches. A lightweight flow detection mechanism is proposed for TG in wireless DCNs in \cite{LightFD} where the detector is designed as a module installed on the Virtual-Switch/Hypervisor. Emulation results show that the detection speed can have a significant impact on system throughput. In our previous work \cite{Celik18design}, we formulated the optimal TG problem and proposed 3-step grooming policy for MFs. This paper extends \cite{Celik18design} by accounting for mission-critical flows and introducing delay analysis along with extensive emulation results. 

\subsection{Main Contributions}
The main contributions of the paper can be summarized as follows:

\begin{itemize}
\item[\checkmark]  
We formulate an optimal TG problem for mission-critical DCNs which mainly consists of three NP-Hard subproblems: 1) Virtual topology design, i.e., lightpath provisioning and routing over the physical topology; 2) Wavelength assignment to the lightpaths; and 3) Developing a grooming policy and routing the groomed traffic requests on the virtual topology.  Although solving such a problem is highly time-consuming even for small-scale DCNs, optimal TG formulation is a key to understand the grooming nature of DCNs and to develop effective heuristic methods.

\item[\checkmark]  
We propose a fast yet high-performance sub-optimal TG policy where CFs and MFs are groomed at servers and ESs to obtain larger rack-to-rack (R2R) mission-critical flows (CFs) and mice flows (MFs), respectively. In order to forward R2R flows, we establish dedicated R2R lightpaths and set their capacity according to delay constraints. These lightpaths are pre-determined and always active to immediately serve any CF and MF arrival. On the other hand, elephant flows (EFs) are forwarded over dedicated S2S express lightpaths whose routes and capacity are dynamically determined based on wavelength and capacity available after CFs and MFs.

\item[\checkmark]  
 CFs are prioritized by considering priority queues where high priority queues are modeled as $M/G/1$ while low priority queues are modeled as $M^a/G/1$ queuing discipline.  Based on queuing theory, we finally provide performance analysis of mission-critical DCNs including waiting times, delay, maximum hop-count, and the probability of blocking. 
 \end{itemize}

\subsection{Paper Organization}

The remainder of the paper is organized as follows: Section \ref{sec:net} presents the network model. Section \ref{sec:opt_tg} formulates the optimal TG problem and Section \ref{sec:tg} develops the proposed sub-optimal solution. Section \ref{sec:perf} analyzes the performance of mission-critical DCNs. Emulation results are presented in Section \ref{sec:res} and Section \ref{sec:conc} concludes the paper with a few remarks.

\begin{figure}[htbp!]
\begin{center}
\includegraphics[width=1\columnwidth]{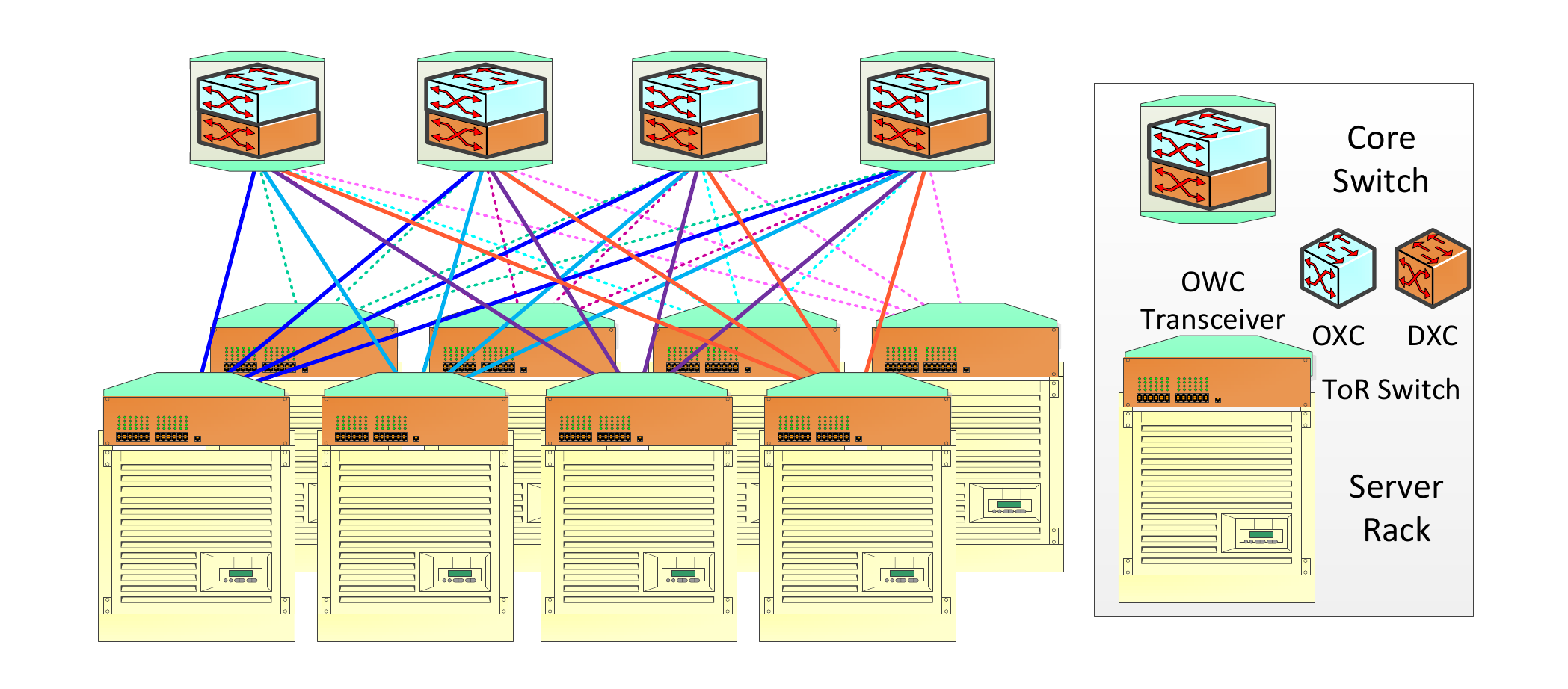}
\caption{Proposed topology for $N=4$ and spine/leaf ratio of $\eta=1/2$.}
\label{fig:top}
\end{center}
\end{figure}

\section{Network Model}
\label{sec:net}

\subsection{Network Topology}   
\label{sec:top}
To employ FSO in DCNs, we adopt a two-layer spine \& leaf Clos architecture where every bottom-layer (leaf-layer) switch is connected to each of the top-layer (spine-layer) switches in a full-mesh topology \cite{Alizadeh2013}. The leaf layer consists of $N$ top-of-rack switches, a.k.a., edge switches (ESs), which is connected to $S$ servers within racks. There also exists $\eta N$, where $\eta$ is the spine \& leaf ratio, core switches (CSs) which are responsible for interconnecting all leaf switches such that every ES connects to every CS. Unlike the spanning tree protocol-based routing within the traditional three-layered architecture, the leaf \& spine architecture employs layer-3 routing which permits all connections to be utilized at the same time while avoiding loops and remaining stable. Furthermore, it is easy to expand the DCN with additional hardware and capacity if oversubscription of links occurs. One of the main concern with the spine \& leaf architecture is the extra cabling complexity which can be solved by the use of WDM-FSO links as follows:

Different from the physical topology designs realized by mechanical devices, reflectors on side walls, and ceiling mirrors, FSO links between the switches can be directly established by an array of optical transceivers as shown in Fig. 1. In this way, performance degradation caused by steering delay and reflection loses can be taken out of consideration. Fig. 1 can be implemented by equipping OWC transceiver box of ESs with a metallic breadboard at their top where transceivers can be fixed such that laser-diodes (transmitters) are fixed on the breadboard via a mount which allows vertical and horizontal alignment towards photo-diodes (receivers) and vice versa. CSs can be located above such that OWC transceiver box and metallic board face downward toward ESs. Alternatively, ESs and CSs can stay at the same level whereas metallic breadboard of CSs is fixed above. 

A similar physical topology set up is also designed in \cite{Arnon2015next} where racks are arranged in circular cells such that neighboring racks can communicate using LoS links and assumes edge switches can communicate with core switches located at higher layers. OWcell is another potential physical topology concept where racks are positioned in regular polygonal topologies to be interconnected with fixed LoS OWC links \cite{Hamza2017}. It is worth noting for our case that as the DCN sizes increases, deployment optimization is necessary to avoid laser beam collisions. Since our main contribution is not focused on optimizing the physical topology of optical wireless DCNs, we assume that rack deployment and laser mount directions are set properly to avoid any laser beam collisions.

\subsection{Hybrid Crossconnect (HXC) Architecture}
\label{sec:switch}
\begin{figure*}[t]
\centering
\includegraphics[width=2\columnwidth]{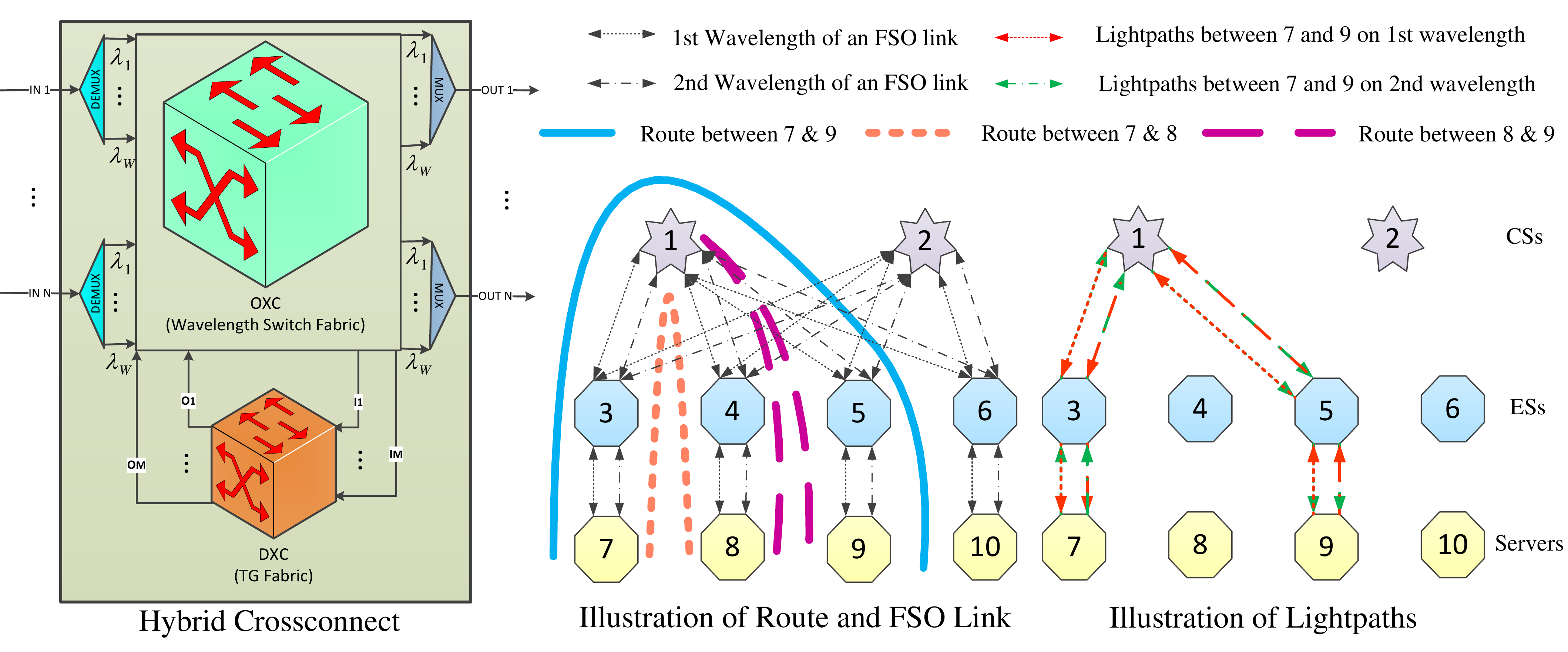}
\caption{Illustration of considered HXC and lightpaths.}
\label{fig:hxc}
\end{figure*}

As shown in Fig. \ref{fig:hxc}, ESs (CSs) are designed as HXCs and have $N$ ($\eta N$) input and output ports which are connected to receivers and transmitters, respectively. The signals at each of the $N$ input ports are first demultiplexed into $W$ individual wavelengths and then processed by optical cross-connect (OXC) or digital cross-connect (DXC) units. OXC is a generic optical switch (i.e., can follow any architectural design in \cite{Yoo2013Elastic, Farrington:2010:Helios, Wang:2010:cThrough,Fiorani2014energy}) and responsible for wavelength routing operations. OXCs may also execute some grooming functions using optical couplers and decouplers \cite{Sankaran2016optical}.

On the other hand, DXCs are traditional packet switches and provide flexibility and bandwidth efficiency by grooming several low-speed flows into a high-capacity lightpath which is simply a wavelength circuit/path between two nodes. It operates on O-E-O conversion principle and can handle $M$ incoming flows at a given time, which determines grooming ratio of an HXC, i.e., $\frac{M}{N}$, $M \leq N$. We assume that servers and DXC are tunable to any wavelength. Notice that DXC ports are also limited by processing speed limitation which is typically in the order of Gbps.  All incoming flows to any HXC port is first handled by OXC based on the flow classifications: EFs are directly routed to the corresponding output ports according to the destination address whereas MFs are first forwarded to the DXC to be groomed by other MFs sharing the same destination, then groomed flow is fed back to the OXC and routed to the final destination.
\subsection{Channel Model}
\label{sec:chmod}
Consider WDM-FSO links formed by directed laser-diode transmitters and photo-diodes receivers which employ intensity-modulation and direct-detection (IM-DD). Thanks to the WDM, FSO links can be treated as parallel channels where received signal at HXC$_m$ from HXC$_n$ on wavelength $\omega$ can be given as 
\begin{equation}
\label{eq:sig}
r_{m,n}^\omega= h_m^n s_{m,n}^\omega + z_m^n, \hspace{10pt} 1\leq \omega\leq W
\end{equation} 
where $s_{m,n}^\omega \in \mathbb{R}_+$ is the transmit intensity, $h_m^n \in \mathbb{R}_+$ represents the optical channel gain, $z_m^n$ is the Gaussian noise  with zero mean and unit variance, and $r_{m,n}^\omega $ is the received signal. Due to hardware and safety concerns, transmit signal intensity has to satisfy an individual and total average intensity constraint given by $\mathbb{E} [s_{m,n}^\omega]=\mathrm{E}_{m,n}^\omega \leq \mathrm{E}$ and $\sum_{\omega} \mathrm{E}_{m,n}^\omega \leq \mathrm{E}_T$, respectively. As optical channel variations are very slow in comparison to the symbol duration \cite{Chaaban2017fundamental}, $h_m^n$ is further assumed to be constant throughout a transmission block and modeled as follows \cite{Kahn1997infrared}. 
If the emitted light-beam from the source is focused on the receiver such that the diverging lens expands the beam to a planar beam diameter, the channel gain $h_m^n$ can be modeled as \cite[Chapter~1]{kaushal2017free} 
\begin{equation}
\label{eq:ch}
h_m^n=\left(G_T \eta_T \eta_{TP}\right)  \left( \frac{\lambda}{4 \pi \delta(m,n)}\right)^2 \left(G_R \eta_R \eta_{\lambda}\right)
\end{equation}
where $G_T=\frac{4 D_T}{\lambda}$ ($G_R=\frac{4 D_R}{\lambda}$) is the transmitter (receiver) gain, $D_T$ ($D_R$) is the transmitter lens (planar beam) diameter, $\lambda$ is the carrier wavelength, $\eta_T$ ($\eta_R$) transmitter (receiver) efficiency, $\eta_{TP}$ is the pointing loss factor, $\eta_{\lambda}$ narrowband filter transmission factor, $\left( \frac{\lambda}{4 \pi \delta(m,n)}\right)^2$ is the free-space path loss, and $\delta(m,n)$ is the distance between transceivers. 
 
Optical intensity channel capacities are studied in \cite{Lapidoth2009capacity}, where upper and lower bounds are shown to converge at high-SNRs as follows
\begin{equation}
\label{eq:cap}
C_{mn}^\omega(E_{m,n}^\omega)= B\frac{1}{2} \log \left(1+\frac{e (h_m^n)^2 (E_{m,n}^\omega)^2}{2 \pi} \right)
\end{equation} 
where $B$ is defined as the bandwidth of a single wavelength. Albeit improved capacity and high fan-out advantage thanks to increased number of wavelengths, employing WDM equipment yields additional expenses such as extra space, monetary costs, power consumption, etc. These expenses are expected to proportionally increase with the required number of wavelengths, $W$, and thus the type of WDM modules employed. For example, coarse WDM modules can support up to 8/16 wavelengths \cite{sector2003spectral} whereas dense WDM modules can provide up to 40/80/160 wavelengths at the expense of higher costs and power consumption \cite{sector2006spectral}. Since WDM multiplexes wavelengths into a single carrier wavelength, the number, and cost of required LDs is directly determined by scaling parameters $\eta$ and $N$, regardless of $W$.

\section{TG Problem Statement and Formulation}
\label{sec:opt_tg}
\subsection{Preliminaries and Problem Statement}
\label{sec:prelim}
In this section, we provide preliminaries to the TG formulation which adopts the following topology definitions:

\begin{mydef}{\textbf{Physical Topology}} $\mathcal{G}_P\left( \mathcal{V},\mathcal{E}_p\right)$ refers to a weighted unidirectional graph where $\mathcal{V}$ and $\mathcal{E}_p$ represent the sets of network nodes and physical links, respectively. A link is defined as a unidirectional physical connection between  nodes, i.e., uplink (UL) or downlink (DL), and weighted by channel gain as given in (\ref{eq:ch}). Each direction consists of $W$ wavelengths and established by a pair of FSO transmitter and receiver. 
\end{mydef}

\begin{mydef}{\textbf{Virtual Topology}}  $\mathcal{G}_V\left( \mathcal{V},\mathcal{E}_
\nu \right)$ corresponds to a weighted unidirectional graph where $\mathcal{V}$ and $\mathcal{E}_\nu$ represent the sets of network nodes and lightpaths, respectively. A lightpath between two generic nodes $i$ and $j$ is denoted as $\mathcal{L}_{i,j}$ and characterized by a route of physical links, a single wavelength of such links, and intensities/capacities allocated to each link.
\end{mydef}

For a visual explanation, let us focus on a simple DCN with two wavelengths as shown in Fig. \ref{fig:hxc}. Physical topology is depicted in black colored dashed and dot-dashed lines for first and second wavelengths, respectively. Example routes are given as $7 \rightarrow 8=\{7,3,1,4,8\}$, $8 \rightarrow 9=\{8,4,1,5,9\}$, and $7 \rightarrow 9=\{7,3,1,5,9\}$ and drawn in different colors and line styles. Note that we can define 4 distinct lightpaths on each of these routes as FSO links are directional and each direction has two wavelengths. A connection demand between two nodes can be met by a combination of the lightpaths. For example, the figure at the right-hand side shows the possible 4 lightpaths for connection requests between 7 and 9. Moreover, these connection requests can alternatively be satisfied by following a different path, e.g., $\{7,3,2,5,9\}$. However, a wavelength of a physical link cannot be exploited by two different lightpaths due to the collision constraint. Throughout the paper, intermediate routing is not allowed, for instance, $\{7,3,2,4,1,5,9\}$ is not a valid routing. Even if such an intermediate routing is allowed, it cannot be an optimal solution for two reasons: 1) It unnecessarily exhausts the computational power of multiple switches along the path and 2) It yields more latency due the processing time of multiple switches along the path.

\begin{table*}[]
\centering
\caption{Notations, parameters, and variables}
\label{tab:given}
\begin{tabularx}{1\textwidth}{l X}
\multicolumn{2}{l}{\textbf{Notations and given parameters:}}                                 \\
\hline
\\
$M_i$ & Number of DXC I/O ports of node $i$.\\
$\bar{S}$     & Total number of servers. \\
$(m,n)$     & Originating and terminating points of a physical (i.e., FSO) link. \\
$(i,j)$     & Originating and terminating points of a lightpath which may traverse multiple FSO links. \\
$t$     &  A flow among a total number of $T$ traffic requests from all sources. Each $t$ corresponds $f^{th}$ flow for a source and destination server pair, i.e., $t\triangleq (s,d,f)$. Note that $t$ may traverse through single or multiple lightpaths, e.g., $t:s\rightarrow (i,j) \rightarrow d$ or $t:s \rightarrow (i,j)\rightarrow(k,l) \rightarrow d$. \\
$F_{mn}$     & Number of FSO links from $m$ to $n$, $F_{mn} \in \{0,1\}$. \\
$W$     & Number of wavelengths on $(m,n)$ iff $F_{m,n}=1$. \\
$C_{mn}^\omega$     &Capacity of $(m,n)$ on wavelength $\omega \in [1,W]$. \\
$\mathcal{D}$  & Bandwidth demand matrix, $\mathcal{D} \in \mathbb{R}_+^{\bar{S} \times \bar{S}}$, where each entry is a vector  of flow requests, i.e., $\mathcal{D}{[s,d]}=\{D_{sd}^f \vert 1\leq f \leq T_s^d\}$ and $T_s^d$ is the total number of flows from $s$ to $d$, $\sum_s T_s^d=T$. Note that $\mathcal{D}$ is necessarily not a symmetric matrix.  \\
$\text{DXC}_i$  & Digital processing capacity of $i$. \\
\multicolumn{2}{l}{\textbf{Optimization Variables:}}                                 \\
\hline
\\
$P_{mn}^{ij,\omega}$     & A binary variable for routing on physical topology; equals 1 iff a lightpath between node pair $(i,j)$ is routed on FSO link $(m,n)$ on wavelength $\omega$. \\
$L_{ij}^{\omega}$    & Number of lightpath from $i$ to $j$ on wavelength $\omega$. \\
$L_{ij}$    & Total number of lightpath from $i$ to $j$, $L_{ij}=\sum_\omega L_{ij}^{\omega}$. Note that $L_{ij}$ and $L_{ji}$ are different variables.\\
$R_{mn}^{ij,\omega \ell}$ & A binary variable to define different physical routes of lightpaths between the same pair of nodes and on the same wavelengths; equals 1 iff a $\ell^{th}$, $1 \leq \ell \leq L_{ij}^{\omega}$, lightpath between node pair $(i,j)$ is routed on FSO link $(m,n)$ on wavelength $\omega$\\
$X_{ij}^{t,\omega\ell}$    & A binary variable, $X_{ij}^{t,\omega\ell}=1$ iff $t=(s,d,f)$ employs $\ell^{th}$ lightpath from $i$ to $j$ on wavelength $\omega$ as an intermediate virtual link.\\
$Y_{ij}^{t}$    & Real valued capacity exploited by $t$ on lightpath(s) between from $i$ to $j$, $Y_{ij}^{t} \geq 0$. \\
$G_{ij}^{tt',\omega \ell}$    & A binary indicator for TG; is $1$ iff $t$ and $t'$ are groomed into lightpath $\ell$ on wavelength $\omega$ from $i$ to $j$. 
\end{tabularx}
\hrule
\end{table*}
TG is usually split into three joint subproblems: 1) designing the virtual topology, i.e., routing and lightpath provisioning over the physical topology; 2) assigning the wavelength to the lightpaths, and 3) developing a grooming policy and routing the groomed traffic requests on the virtual topology. Noting that each of these sub-problems is NP-hard  \cite{Zhur2002mesh}, TG is also an NP-hard problem which falls into mixed-integer linear programming (MILP) class. Considering the resulting high number of binary and integer variables, obtaining an optimal TG solution is highly time-consuming even for a small size of DCNs. However, we believe formulating the optimal problem is necessary to gain an insight into the inherent features of TG within DCNs and to develop fast yet high-performance heuristic solutions.     
\subsection{Problem Formulation}
\label{sec:formula}
Using standardized TG formulations \cite{Zhur2002mesh, Mustafa2006design}, we formulate this MILP problem based on parameters/variables in Table \ref{tab:given}. The main design goal can be set to any objective function, e.g., throughput maximization,  delay minimization, the minimum number of required wavelength, minimum power consumption, etc. Unlike the traditional TG in passive optical mesh networks, we also need to take the intensity allocation of the links as they do not have fixed capacities as in fibers. Accordingly, we formulate the problem based on the following assumptions and constraints:

\begin{enumerate}
\item[$\mathrm{AS}_1:$]  HXCs are not capable of wavelength conversion. Thus, a lightpath must be routed on the same wavelength.
\item[$\mathrm{AS}_2:$]  Bifurcation of flows is not allowed. That is, a connection request cannot be divided and routed separately.
\item[$\mathrm{AS}_3:$]  DXCs can groom as many flows into a lightpath as needed, as long as DXC processing capability and wavelength capacity are not exceeded.
\item[$\mathrm{AS}_4:$]  I/O ports of servers and DXCs are tunable to any wavelength. 
\end{enumerate}

$\blacklozenge$ \textit{\textbf{Virtual topology constraints:}}
\begin{align}
\label{eq:vtc1}
\sum_\omega \sum_{j, j\neq i} L_{ij}^{\omega} & \leq M_i, \forall i \\
\label{eq:vtc2}
\sum_\omega \sum_{j, j\neq i} L_{ji}^{\omega} & \leq M_i, \forall i \\
\label{eq:vtc3}
\sum_\omega L_{ij}^\omega& =L_{ij}, \forall i ,j
\end{align}
where (\ref{eq:vtc1}) and (\ref{eq:vtc2}) ensure that number of originating and terminating lightpaths at HXC$_i$ do not exceed the number of DXC I/O ports, respectively. That is, (\ref{eq:vtc1}) and (\ref{eq:vtc2}) limits the total number of lightpaths which can be processed by the DXC  at a given time. On the other hand,  (\ref{eq:vtc3}) is the expression for the total number of lightpaths provisioned between nodes $i$ and $j$, which is exploited later by other constraints. 

$\blacklozenge$ \textit{\textbf{Lightpath routing (flow conservation) constraints:}}
\begin{align}
\label{eq:fcc1}
\sum_m P_{mi}^{ij,\omega}F_{mi}&=\sum_n P_{jn}^{ij,\omega}F_{jn}=0, \forall i,j,\omega\\
\label{eq:fcc2}
\sum_m P_{mj}^{ij,\omega}F_{mj}&=\sum_n P_{in}^{ij,\omega}F_{in}=L_{ij}^\omega, \forall i,j,\omega\\
\label{eq:fcc3}
\sum_m P_{mk}^{ij,\omega}F_{mk}&-\sum_n P_{kn}^{ij,\omega}F_{kn}=0, \forall i,j,k,k\notin \{i,j\} \\
\label{eq:fcc4}
\sum_{i,j} P_{mn}^{ij,\omega} & \leq F_{mn}, \forall m,n,\omega\\
\label{eq:fcc5}
\sum_{i,j} R_{mn}^{ij,\omega\ell} & \leq P_{mn}^{ij,\omega}, \forall m,n,\omega,\ell\\
\label{eq:fcc6}
\sum_{m,n,i,j,\omega} P_{mn}^{ij,\omega} &\sum_{\ell=1}^{L_{ij}^\omega} R_{mn}^{ij,\omega\ell} =1. \\
\label{eq:fcc7}
\sum_{m,n,\omega} P_{mn}^{ij,\omega} &=4, \forall i,j.
\end{align}
where (\ref{eq:fcc1}) assures that there are no incoming (outgoing) flows for originating (terminating) node $i$ ($j$) of lightpath $(i,j)$ on wavelength $\omega$. Supported by the underlying physical topology, the total number of lightpaths on wavelength $\omega$ from $i$ to $j$ is given in (\ref{eq:fcc2}). Constraints in (\ref{eq:fcc3}) and (\ref{eq:fcc4}) ensure the wavelength continuity and protection against lightpath collisions, respectively. Constraints (\ref{eq:fcc5}) and (\ref{eq:fcc6}) permits at most one lightpath to be routed on $(m,n)$ among all lightpaths. Finally, (\ref{eq:fcc7}) prevents intermediate routing by setting total number of physical link hops to $4$ as a valid route (S $\to$ ES $\to$ CS $\to$ ES $\to$ D) takes exactly 4 hops to reach the destination. 

 $\blacklozenge$\textbf{ \textit{Connection topology constraints:}}
 \begin{align}
  \label{eq:ctc0}
 X_{ij}^t&= \sum_{\omega}\sum_{\ell=1}^{L_{ij}^\omega} X_{ij}^{t,\omega\ell}, \forall i,j,t \\
 \label{eq:ctc1}
 \sum_i X_{is}^t&= \sum_j X_{dj}^t=0, \forall t \\
  \label{eq:ctc2}
 \sum_{i,i\neq k} X_{ik}^t &= \sum_{j,j \neq k} X_{kj}^t, \forall t, k, k \notin \{s,d\}
 \end{align}
where (\ref{eq:ctc0}) defines the total number of connections established on all wavelengths and lightpaths from $i$ to $j$. 
%That is some portion of traffic $t$ can be split to some different lightpath and wavelength combinations. 
The constraint in (\ref{eq:ctc1}) guarantees that there is no incoming and outgoing traffic to the source and destination of a traffic request, respectively. On the other hand,  (\ref{eq:ctc2}) preserves the continuity of the flows on single or multiple lightpaths. Even though different flows between a source-destination pair are allowed to be split to different lightpaths, wavelengths, or routes, non-bifurcation keeps a certain flow intact and exploit only one lightpath, wavelength, and physical route tuple. 

$\blacklozenge$ \textit{\textbf{Non-bifurcation constraints:}}
\begin{align}
\label{eq:nbc-1}
X_{ij}^t & \leq 1, \forall i,j,t\\
\label{eq:nbc0}
\sum_{i,j} X_{ij}^t& \leq 1, \forall t\\
\label{eq:nbc1}
G_{ij}^{tt}&=X_{ij}^t, \forall i,j,t\\
\label{eq:nbc2}
G_{ij}^{tt'}&= \sum_{\omega}\sum_{\ell=1}^{L_{ij}^\omega}G_{ij}^{tt', \omega \ell}, \forall i,j,t\neq t'\\
\label{eq:nbc3}
G_{ij}^{tt'}& \leq \rfrac{1}{2}\left(X_{ij}^t+X_{ij}^{t'}\right), \forall i,j,t\neq t' \\
\label{eq:nbc4}
L_{ij}&=G_{ij}^{tt}+\sum_{t', t' \neq t} \left( G_{ij}^{t't'} -\bigvee_{x=1}^{t'-1} G_{ij}^{t'x} \right)
\end{align}
where (\ref{eq:nbc-1})-(\ref{eq:nbc4}) satisfy non-bifurcation of traffic among lightpaths between different nodes, among wavelengths of a lightpath between the same pair of nodes, and among different physical routes a lightpath between the same pair of nodes and on the same wavelength. In other words, by imposing this set of equations flows are enforced to stay as a whole beginning from the source until the destination.

$\blacklozenge$ \textit{\textbf{Capacity and delay constraints:}}
\begin{align}
\label{eq:EC1}
\mathrm{E}_{m,n}^\omega & \leq \mathrm{E} \\
\label{eq:EC2}
\sum_{\omega} \mathrm{E}_{m,n}^\omega &  \leq \mathrm{E}_T\\
\label{eq:nbc5}
C_{mn}^\omega (E_{m,n}^\omega)& \geq \sum_{i,j,t} Y_{ij}^t X_{ij}^{t,\omega\ell} R_{mn}^{ij, \omega \ell}, \forall m,n  \\
\label{eq:nbc6}
\text{DXC}_i& \geq \sum_t \left( \sum_{j, j \neq i} X_{ij}^t Y_{ij}^t + \sum_{j, j \neq i} X_{ji}^t Y_{ji}^t \right), \forall i\\
\end{align}
where (\ref{eq:EC1}) and (\ref{eq:EC2}) constitute the upper bounds on single wavelength intensity and total intensity allocation of an FSO link, respectively. Capacity is a function of the intensity variable $\mathrm{E}_{m,n}^\omega $ as formulated in (\ref{eq:cap}) and constraint in (\ref{eq:nbc5}) ensures that total traffic request of set of flows, which are groomed on the same physical route on a certain lightpath and wavelength pair, must comply with the capacity of that route, i.e., the lowest capacity along the physical route. Finally, the total amount of incoming and outgoing traffic to be processed is limited by processing capability of nodes as in (\ref{eq:nbc6}). Please note that the bandwidth demand of each flow is determined based on flow size and maximum affordable delay of the flow, which satisfies the delay constraints. 
\section{TG Policy Design for DCNs}
\label{sec:tg}
In this section, we develop a TG policy for delay constrained DCNs where flows are classified into EFs, MFs, and CFs. While CFs and MFs have high priority, EFs have low priority. The maximum delays affordable by high and low priority flows are denoted as $\tau_h$ and $\tau_\ell$, respectively. The delay constraint is essential especially for the mission-critical applications as delayed packets are considered useless. Accordingly, we define the following TG policy rule-set for the proposed model.

\begin{enumerate}
\item[$\mathrm{PS}_1$] 
We assume that flows can be classified, via packet classification or flow matching, in a timely and efficient manner. While MF and CFs are groomed into larger traffic, EFs are not groomed and treated separately. 
\item[$\mathrm{PS}_2$]  
Lightpaths are first provisioned for groomed CFs and then groomed MF. The residual network resources are then used to provision EF lightpaths.
\item[$\mathrm{PS}_3$]  
Bifurcation at HXC's are not allowed as splitting and combining EFs can consume a significant portion of DXC processing capability at the origin and terminal points of lightpaths, respectively. On the other hand, splitting CFs and MF may not be necessarily efficient. 
\item[$\mathrm{PS}_4$]  
Aside from $W$ wavelengths, there exists a dedicated wavelength for a central controller to broadcast signals for TG and intensity allocation commands. Current intensity and wavelength availability state of DCN is formulated in graphs $\mathcal{G}_I\left(\mathcal{V},\mathcal{I}\right)$ and $\mathcal{G}_A \left(\mathcal{V},\mathcal{A} \right)$, respectively, where $\mathcal{V}$ is the set of nodes, $\mathcal{I}$ presents available light intensity, and $\mathcal{A}$ presents binary edge weights for wavelength availability. These graphs are always kept updated over the control wavelength.
\end{enumerate}

\subsection{TG and Lightpath provisioning for MF}
CF and MF grooming is designed to take place at source and switches in three TG steps as follows: 
\begin{enumerate}
\item \textit{S2S Step:} 
Each server grooms all flow arrivals destined to the same destination server. Indeed, the web and database servers receive a very large number of concurrent requests where the clients could be from the same subnet or different subnets. Since servers know whatever packets that belong to which flow and their destination subnet, handling flows destined to the same server is helpful to reduce workload and complexity of dealing with individual flow entities.  

\item \textit{S2R Step:} 
Each server further grooms all S2S flows destined to the same destination and transfer it to the ESs. The packets of the same flow are groomed into a jumbo packet (i.e., frame) and propagated to the ES or the virtual switch if the source and destination reside on the same machine. Enabling jumbo frames can improve the efficiency of data transmissions, and hence, the network performance. The processing overhead will be reduced because the CPUs of DXCs need to process a single jumbo packet at a time rather than multiple packets. Therefore, the main motivation behind the S2S and S2R steps is to reduce the workload on digital switches as they have limited processing speed at each port. 

\item  \textit{R2R Step:} 
Received S2R flows are then groomed according to their destination rack to obtain R2R flows. At this point, R2R flows are only needed to be routed toward their final destination rack without being processed by core switches, which allows employing very fast and effective routing mechanisms. 
\end{enumerate}

Above steps are applied to CFs and MF separately as they have a different priority and delay requirement. Thanks to WDM, a large set of route and wavelength possibility is already available. Hence, groomed CFs (MFs) from Rack$_i$ to Rack$_j$ is provided with a dedicated lightpath which is denoted as $\mathcal{L}_{i,j}^c$ ($\mathcal{L}_{i,j}^m$) and defined by route, wavelength, and intensity allocation tuple.  Notice that the grooming and de-grooming occur at the edge switches of source and destination servers, respectively. On the other hand, core switches function as a router without involving in any data processing task. Thus, packets stay intact and are routed all together along the predetermined lightpaths. Moreover, since we define only one route for each R2R pair, it is impossible to forward portions of the groomed R2R flows over multiple paths which are never defined. Unlike the wired DCNs with the uniform link capacities, wireless DCNs experience heterogeneous link capacities due to the distance differences. Thus, to determine R2R lightpaths, we first create matrices $\vect{K}$ and $\vect{D}$ to record the number of short paths and corresponding total intensity cost between rack pairs, respectively. After that, lightpath provisioning starts from rack pairs with a lower number of shortest paths. In this way, route diversity of rack pair $(i,j)$ is not affected by rack pair $(k,l)$ if $K_{ij}<K_{kl}$. Among pairs with the same number of shortest path, the tie is broken by prioritizing the pair with the highest cost. This iterative procedure is repeated until all lightpaths are determined. Required minimum number of wavelengths to set dedicated R2R lightpaths can be derived as 
\begin{equation}
W \geq \left \lceil \frac{2 (N-1)}{\eta N} \right \rceil.
\end{equation}
which is exactly based on the fact that intermediate routing is not allowed and a pair of wavelength is needed for each rack pairs, one for MF and the other for EF. Since R2R flows are assigned to dedicated lightpaths and predetermined route-wavelength pairs, the proposed approach has low complexity and incurs almost no decision making delay. Unlike the generic topology is shown in Fig. \ref{fig:top}, CSs are not required to equipped with DXCs as TG is executed only by servers and ESs in the proposed TG policy. Since R2R flow size is limited by DXC port speed, the capacity of MF wavelengths must also be upper bounded by DXC speed, which naturally opens some room for extra intensity required by some EFs as explained next.

\begin{algorithm}[t!]
\footnotesize
 \caption{Proposed TG Approach}
  \label{alg:TG}
\begin{algorithmic}[1]
 \renewcommand{\algorithmicrequire}{\textbf{Input:}}
 \renewcommand{\algorithmicensure}{\textbf{Output:}}
\Require $\mathcal{G}_P\left( \mathcal{V},\mathcal{E}_p\right)$, $W$ 
 \State $\mathcal{G}_V\left( \mathcal{V},\mathcal{E}_
\nu \right) \gets $ Initialize the virtual topology
 \State $ \mathcal{G}_I\left(\mathcal{V},\mathcal{I}\right) \gets$ Initialize the residual intensity graph
 \State $\mathcal{G}_A \left(\mathcal{V},\mathcal{A} \right) \gets $  Initialize the wavelength availability graph
 \State  $(\mathcal{L}_{i,j}^c,\mathcal{G}_V, \mathcal{G}_I,\mathcal{G}_A ) \gets \textsc{Lightpath Provisioning ($\lambda_{i,j}^c$)}$
  \State  $(\mathcal{L}_{i,j}^m,\mathcal{G}_V, \mathcal{G}_I,\mathcal{G}_A ) \gets \textsc{Lightpath Provisioning ($\lambda_{i,j}^m$)}$
\For{each traffic request arrival $t$}
\If {$t$ is an CF}
\State Employ 3-Step TG, i.e., S2S, S2R, and R2R. 
\State Forward groomed traffic over  R2R-CF lightpaths
\ElsIf {$t$ is an MF}
\State Employ 3-Step TG, i.e., S2S, S2R, and R2R. 
\State Forward groomed traffic over  R2R-MF lightpaths.
\Else
\State $\mathcal{L}_{s,d}^t \gets \textsc{EF Lightpath Provisioning ($t$)}$
 \State $\mathcal{G}_V\left( \mathcal{V},\mathcal{E}_\nu \right) \leftarrow $ Update the virtual topology
 \State $\mathcal{G}_I\left(\mathcal{V},\mathcal{I}\right) \leftarrow $ Update residual intensity
 \State $\mathcal{G}_A \left(\mathcal{V},\mathcal{A} \right) \leftarrow $  Update wavelength availability
\EndIf
\EndFor
\vspace{2pt}
\hrule 
\vspace{2pt}
\Procedure{Lighpath Provisioning }{$\lambda_{i,j}$}
\State $(\vect{K}, \vect{D}) \gets $ Create the KSP matrix  $\vect{K}$ and corresponding cost matrix $D$.
\While{There is unprovisioned lightpaths}
\State $(i,j) \gets $Determine the rack pair
\State $E_{m,n}^\omega \leftarrow$ Allocate intensity as per (\ref{eq:CFMFint})
\State $\mathcal{L}_{i,j}\leftarrow$ Record lightpath Rack$_i \rightarrow$Rack$_j$.
\State $(\vect{K}, \vect{D}) \gets $ Update $\vect{K}$ and $\vect{D}$.
 \State $\mathcal{G}_V\left( \mathcal{V},\mathcal{E}_
\nu \right) \leftarrow $ Update the virtual topology
 \State $\mathcal{G}_I\left(\mathcal{V},\mathcal{I}\right) \leftarrow $ Update residual intensity
 \State $\mathcal{G}_A \left(\mathcal{V},\mathcal{A} \right) \leftarrow $  Update wavelength availability
 \EndWhile

\hspace{-9pt} \Return $\mathcal{L}_{i,j}$, $\mathcal{G}_V\left( \mathcal{V},\mathcal{E}_\nu \right), \mathcal{G}_I\left(\mathcal{V},\mathcal{I}\right),\mathcal{G}_A \left(\mathcal{V},\mathcal{A} \right) $
\EndProcedure

\vspace{2pt}
\hrule 
\vspace{2pt}
\Procedure{EF Lighpath Provisioning }{$t$}
\State $\mathcal{R}_s^d \gets $ Determine feasible routes
\State $E_{m,n}^{\omega,t} \gets$ Calculate intensities of $\mathcal{R}_s^d $ as per (\ref{eq:avpwr})
\State $C_t^r \gets$ Compute route capacities as per (\ref{eq:Ctr})
\State $r^\star\gets$ Decide on the \textit{Best-Fit} route as per (\ref{eq:rstar})
\State $\omega \gets$ Determine on wavelength \textit{First-Fit} Scheme
\State $\mathcal{L}_{s,d}^t \gets$ Record the lightpath from $s$ to $d$

\hspace{-8pt} \Return $\mathcal{L}_{s,d}^t$,
\EndProcedure

 \end{algorithmic}
 \end{algorithm}

\subsection{Intensity Allocation}
\label{sec:allocation}
 WDM based FSO links provide two significant advantages: increased channel diversity (a.k.a, high fan-out) and capacity flexibility. While assigning wavelengths with the fixed intensities/capacities (as in the WDM fiber links) ignores the flexibility provided by optical wireless technology, allocating unnecessarily high intensities to a certain wavelength destroys the wavelength availability for future flows.\footnote{For example, consider an extreme scenario where a single wavelength is allocated with the full intensity and remaining wavelengths cannot be available at all. Alternatively, one can consider a fixed-uniform intensity allocation scheme which does not account for traffic characteristics of flows. In such a case, bandwidth utilization can reduce in a significant manner since some wavelengths are allocated with more/less intensity than they require.} To have an improved bandwidth utilization, it is necessary to have a fast yet efficient intensity allocation method for adjusting link capacities according to groomed traffic requirements. 

We first allocate intensities of the link-wavelength pairs along the R2R CF (MF) routes as follows: Average groomed CF (MF) size between Rack$_i$ and Rack$_j$ is denoted as $\kappa_{i,j}^c=\varpi \lambda_{i,j}^c$ ($\kappa_{i,j}^m=\varpi \lambda_{i,j}^m$) where $\varpi$ is the flow size and $\lambda_{i,j}^c$ ($\lambda_{i,j}^m$) is the total CF (MF) arrival rate from servers of Rack$_i$ to servers of Rack$_j$. Accordingly, if a groomed R2R traffic is routed over an FSO link between nodes $m$ and $n$, required transmission intensity on this link can be calculated from (\ref{eq:cap}) as follows
\begin{equation}
\label{eq:CFMFint}
E_{mn}^\omega = \sqrt{\frac{2 \pi \left(2^\frac{2 \kappa_{i,j}^{k}}{B \tau_h} -1\right)}{e (h_m^n)^2}}, k\in \{c,m\}
\end{equation}
where $\tau_h$ is the affordable maximum delay by CFs and MFs. Available wavelengths and intensity levels after CF/MF allocations are shared among EFs based on a fair-share policy which guarantees $\frac{E_T}{W}$ intensity allocation for each wavelength. Thus, the required intensity by traffic request $t$ of size $\kappa_t$ on wavelength $\omega$ of link $(m,n)$ is given as
\begin{equation}
\label{eq:avpwr}
E_{mn}^{\omega t}= \min \left \{E_{mn}^a-(W_{mn}^a-1)\frac{E_T}{W}, \sqrt{\frac{2 \pi \left(2^\frac{2 \kappa_t}{B \tau_e} -1\right)}{e (h_m^n)^2}} \right \}
\end{equation}  
where $E_{mn}^a$ is the available intensity on $(m,n)$ after CF/MF allocations, $W_{mn}^a=\vert \mathcal{W}_{mn}^a\vert$ is available number of wavelengths for EFs on  $(m,n)$, and $ \mathcal{W}_{mn}^a$ is the set of these wavelengths. In (\ref{eq:avpwr}), the first term is the total intensity available after CF/MF allocations and fair-share policy warranty whereas the second term is the demanded intensity to meet requirements of flow $t$. Therefore, \eqref{eq:avpwr} enforces each wavelength to exploit the exact demanded intensity as long as it obeys fair-share policy. That is, a power demand exceeding equalized power policy can be obtained from the room opened by less demanding flows. 

\subsection{Lightpath provisioning for EFs}
As aforementioned, EFs are treated separately and are not subject to TG as they already have significant traffic requests. Since DXCs have limited processing power which constitutes a network bottleneck, forwarding EFs e along with the MFs may cause unnecessarily extra processing overhead. Hence, once an EF is detected, an S2S lightpath is required to be established between the source and destination servers, which is routed over OXCs and terminated after the session completion. That is, EFs are sent express through OXCs on a specific wavelength and route pair. Based on $\mathcal{G}_I$ and $\mathcal{G}_A$, each server maintains shortest-paths lists toward all destinations, available total light intensity on each hop of shortest-paths, and indices of available wavelengths. 

Let us consider a traffic request classified as an EF of size $\kappa_t$. Taking the wavelength continuity into consideration, a route is feasible only if there exists an available wavelength at its every hop. Denoting the set of such routes from source $s$ to destination $d$ as $\mathcal{R}_s^d$, the achievable capacity of the routes are determined as follows 
\begin{equation}
\label{eq:Ctr}
C_{t}^{r}=\min_{\substack{(m,n) \in r}} C_{mn}^\omega(E_{mn}^{\omega t}), \hspace{10pt} r \in \mathcal{R}_s^d
\end{equation}
which determines the route capacity by the minimum capacity of the links along the routing path.
Based on calculated route capacities and available number of wavelengths, route is determined based on a \textit{best-fit} policy 
\begin{equation}
\label{eq:rstar}
r_{t}^\star=\argmax_{ r \in \mathcal{R}_s^d} \left(W_{t}^r C_{t}^{r} \right),
\end{equation}
where $W_{t}^r=\vert \bigcap_{\substack{(m,n) \in r}}\mathcal{W}_{mn}^a\vert ,  r \in \mathcal{R}_s^d,  (s,d) \in t$. Apparently, (\ref{eq:rstar}) favors for routes with higher available wavelength and capacity. The next step is assigning a wavelength to the selected route, which can be done by a variety of methods, e.g., random, first-fit, least/most used, etc. Since they are shown to perform very close to each other  \cite{zang2000review}, we employ the first-fit scheme and assign the lowest index of available wavelengths. Please note that route selection in (\ref{eq:rstar}) is desirable due to its low computational complexity.
 
If a flow is rejected on a certain route, it competes for the second best-fit route calculated in (\ref{eq:rstar}), and so forth.  In the case of multiple servers decide to use the same wavelength for their EFs, the centralized scheduler follows the Shortest Job First (SJF) scheduling policy. Accordingly, competing flows are ordered based on their sizes and the flow with the smallest size is served first. The packets of the other flows are kept wait on a low priority queue until the wavelength is available to serve the second priority flow. Contingent upon all above details, proposed TG policy is summarized in Algorithm \ref{alg:TG}. In the realm of optical mesh networks, traffic grooming (TG) is a traditional way of effectively utilizing the available bandwidth by combining sub-wavelength capacity flows into larger flows. On the contrary, proposed Algorithm 1 goes beyond of this traditional approach by proposing a simple yet expeditious three steps grooming method developed especially for DCN architecture. Thanks to the flexibility of optical wireless links, efficient bandwidth utilization can be obtained by adjusting the wavelength capacities via power/intensity allocation based on traffic load characteristics of DCNs. Therefore, Algorithm 1 applies to the DCNs with adjustable and heterogeneous link capacities as well as to DCNs with fixed-uniform link capacities (e.g., fiber links).

\section{Delay Analysis of Mission-Critical DCNs} 
\label{sec:perf}

In this section, our objective is to obtain an approximate expression of the blocking probability due to either buffer overflow or violating certain end-to-end delay threshold. We consider a typical mission-critical application servers scenario where $S$ servers receive specific information, and then, transmit them over multiple switches to another application server for processing and decision making. 

\subsection{Traffic Model}\label{sec:model}

Data packet arrival is modeled as an exponential inter-arrival times with rates $\lambda_s, s=1,2,...,S$. The intermediate FSO switches are modeled as single server facilities with two priority queues and are responsible for forwarding the data to the application server. Arrival rates of the MF packets are denoted as $\lambda_M$ and scheduled to the higher priority queue denoted as $Q^H_k$, where $k=1,2,...,K$ is the switch index.  On the other hand, the arrival rate of EF packets is denoted as $\lambda_E$ and scheduled to the lower priority queue $Q^L_k, k=1,2,...,K$. In some special cases, some non-critical packets could suddenly become critical according to the current context. Therefore, such packets will be scheduled in the high-priority queue with probability $1-\flat$. In general, both critical and non-critical packet arrivals are assumed to be Poisson distributed with rates $\lambda_E$ and $\lambda_M= \sum_{s=1}^S \lambda_s$, respectively.

Since there are two different types of traffic and consequently priority queues, the total rate of high-priority traffic $\lambda_h$ and low-priority traffic $\lambda_\ell$ is given by:
\begin{eqnarray}
    \lambda_h&=&\lambda_m+\lambda_c =\lambda_m + \left(1-\flat\right)\cdot\lambda_E \\
    \lambda_\ell&=& \flat \cdot \lambda_E
\end{eqnarray}
where $\lambda_c$ can be regarded as the arrival rate of CFs. Also, the size and rate of EFs are generally larger than the size of the MFs as well as CFs. However, the primary challenge with the EFs is their rate rather than their size. In practice, the packets of a data flow arrive at a network switch in a concatenate sets representing the congestion windows (CWND) of their senders. As the elephant flow has a larger data size, its CWND has enough time to grow up faster by harnessing available capacity as well as left spaces of the completed MFs. 
 
 Thereby, the CWND of elephant flows is in order of magnitude larger than the CWND of MFs. In some cases, an MF complete before its CWND examines the maximum available capacity~\cite{routerbuffer, tcpfouad}. Network switches are assumed to operate following exponential distributed service time of mean $\overline{\mathcal{X}}_{k}, \: k=1,2,...,K$~\cite[Chapter 3]{datanetworks}.  To maintain a guaranteed behavior for the mission-critical traffic, the expected delay experienced by each data packet (since its transmission till it reaches its destination) must not exceed a predefined threshold $\mathcal{T}_{\rm{QoS}}$. Therefore, a data packet with a cumulative delay exceeding $\mathcal{T}_{\rm{QoS}}$ upon arrival to the application server will simply be flagged and then an appropriate decision will be employed. Thus, if we let $\mathfrak{D}$ be a random variable that stands for the end-to-end delay, then the probability of dropping a data packet at the application server is given by
\begin{equation}
 \mathcal{P}_{\rm{blocking}}=\rm{Pr}\left\{\mathfrak{D}>\mathcal{T}_{\rm{QoS}}\right\}
\end{equation}

In this case, our objective is to compute the maximum allowable hop-count $\rm{H}^*$ such that $\mathbb{E}[\mathfrak{D}] < \mathcal{T}_{QoS}$, and estimate the packet dropping probability for a given number of hops in the network. Looking into the above objectives, we aim at analyzing how the above metrics vary with service, scheduling, and application policies considering reconfigurable FSO links for our WDCN model. We assume that $\tau$ is the propagation delay between the application servers and the optical switches.

\subsection{Delay Analysis}\label{sec:analysis}
As mentioned earlier, our network consistent with $K$ switches with two priority queues and possibly with different service rate distributions. Initially, we are interested in obtaining the average waiting time $\overline{\mathcal{W}_p}$ and second moment $\overline{\mathcal{W}_p^2}$ for priority $p\in\{\ell,\,h\}$ that will be experienced by sampling data packets at each switch. Since the output of each queue is approximated as a Poisson process, we look into each switch in isolation and temporarily drop the switch index $k$. Thus, the waiting time of a high priority packet is given as 
\begin{equation}\label{WHEq}
\mathcal{W}_{h} = \sum_{i=1}^{\mathcal{N}_h} \mathcal{X}_{i} + \mathcal{R}
\end{equation}
where, $\mathcal{X}_{i}$ is the service time of packet $i$ and $\mathcal{R}$ is the residual time. From (\ref{WHEq}), using Little's formula, we obtain the average waiting time of the high priority queue $\overline{\mathcal{W}}_h$ as,
\begin{equation}\label{WH1stM_1}
\overline{\mathcal{W}}_h = \frac{\overline{\mathcal{R}}}{(1-\rho_h)}
\end{equation} 
where $\rho_h = \lambda_h\cdot \overline{\mathcal{X}}$ is the fraction of time a switch is serving high priority traffic. By raising both sides of \eqref{WHEq} to the second  power and taking expectation, the second moment of waiting time for high-priority traffic is derived as
\begin{align}\label{WH2ndM_1}
\nonumber\overline{\mathcal{W}_h^2} &=\overline{\mathcal{N}}_h \cdot\rm{Var}(\mathcal{X}) + \rm{Var}(\mathcal{R}) + \text{Var}(\mathcal{N}_h)\cdot \overline{\mathcal{X}}^2\\ 
&\approx\overline{\mathcal{N}}_h \cdot\rm{Var}(\mathcal{X}) + \rm{Var}(\mathcal{R}).
\end{align}
where the approximation is based on the assumption of negligible last term as  $\text{Var}(\mathcal{N}_h)$  is typically small especially for MF that have limited traffic. We understand from the above that if high priority traffic is limited, then high priority queues contain at most one packet almost all the time, and waiting time for high priority packets is chiefly due to the residual time only. For lower-priority traffic, the waiting time of a low priority packet can be expressed as,
\begin{equation}\label{wlEq1}
\mathcal{W}_\ell = \sum_{i=1}^{\mathcal{N}_h} \mathcal{X}_{i} + \sum_{i=1}^{\mathcal{N}_\ell} \mathcal{X}_{i} + \sum_{i=1}^{\mathcal{W}_\ell\cdot \lambda_h} \mathcal{X}_{i}+\mathcal{R}
\end{equation}
which essentially implies that the waiting time of a low-priority packet is the summation of four components: 1) The time to serve existing high priority packets, 2) The time to serve existing low-priority packets that are ahead in the queue, 3) The time to serve new high priority packets that arrive while the low-priority packet is waiting, and 4) Residual time due to the service. Accordingly, the average waiting time for the low priority queue denoted by $\overline{\mathcal{W}}_\ell$ is given as\footnote{\eqref{WL1Eq} and \eqref{WL2Eq} are based on the following assumptions: 1) The distribution of interarrival time is exponential and 2) The distribution of network switches service-time is general.}
\begin{equation}\label{WL1Eq}
\overline{\mathcal{W}}_\ell = \frac{\overline{\mathcal{R}}}{(1-\rho_h)\,(1-\rho_h-\rho_\ell)} ,
\end{equation}
and the second moment is derived as
\begin{eqnarray}\label{WL2Eq}
\nonumber \overline {\mathcal{W}_\ell^2} &=& \overline{\mathcal{R}^2} +\vartheta \cdot \text{Var}(\mathcal{X}) +\vartheta^2\cdot \overline{\mathcal{X}}^2+ 2\,\vartheta\,\overline{\mathcal{X}}\cdot \overline{\mathcal{R}} + \theta \cdot \overline{\mathcal{X}^2} \\ 
&\approx& \overline{\mathcal{R}^2} +\vartheta \cdot \text{Var}(\mathcal{X}) + \vartheta^2\cdot \overline{\mathcal{X}}^2+ 2\,\vartheta\,\overline{\mathcal{X}}\cdot \overline{\mathcal{R}} 
\end{eqnarray}
where the coefficient $\vartheta$  is defined as
\begin{equation*}
\vartheta = \overline{\mathcal{N}_\ell} + \overline{\mathcal{N}_h}+\lambda_h\cdot \overline{\mathcal{W}_\ell},
\end{equation*}
and $\theta$ is the variance of the number of packets in the four cases mentioned earlier and is assumed to be negligible in a steady state. Hence the last approximation follows. One immediate check is to note that (\ref{WL2Eq}) reduces to (\ref{WH2ndM_1}) whenever  $\lambda_E\to 0$, which agrees with expectation because in the latter case lower-priority traffic essentially becomes higher-priority traffic.

To complete our analysis, we need to find the first moment and second moment of $\mathcal{R}$ which are essential to calculate the average waiting time for the packets in both the high priority queue and the low priority queues. The first moment of the residual time is defined as 
\begin{equation}\label{MeanREq}
\overline{ \mathcal{R}} = \frac{\overline{n}}{2}\left(\left(\rho_h+\rho_\ell\right)\cdot \frac{\overline{ \mathcal{X}^2}}{\overline{ \mathcal{X}}}\right).
\end{equation}
which is obtained by employing the first moment of the remaining service time of M/G/1 queue, $\overline{ \mathcal{R}} = \frac{{1}}{2}\left(\left(\rho_h+\rho_\ell\right)\cdot \frac{\overline{ \mathcal{X}^2}}{\overline{ \mathcal{X}}}\right)$, and Little's law. Based on the law of total expectations, we have $\overline {\mathcal{R}^2} = (\rho_h+\rho_\ell)\cdot \overline{\mathcal{R}^2}=\frac{{1}}{3} \left( \left(\rho_h+ \rho_\ell \right)\, \frac{\overline{\mathcal{X}^3}}{\overline{\mathcal{X}}} \right)$. Applying Little's law yields
\begin{equation}
\label{2MeanREq}
\overline {\mathcal{R}^2} = (\rho_h+\rho_\ell)\cdot \overline{\mathcal{R}^2}=\frac{\overline{n}^2}{3} \left( \left(\rho_h+ \rho_\ell \right)\, \frac{\overline{\mathcal{X}^3}}{\overline{\mathcal{X}}} \right).
\end{equation}
Accordingly, \eqref{MeanREq} and \eqref{2MeanREq} can be used to obtain $\text{Var}(\mathcal{R}) = \overline{\mathcal{R}^2} - \overline{\mathcal{R}}^2$. 

\subsection{Maximum Hop-count}

The maximum allowable hop-count that respects the QoS delay constraint is given by
\begin{equation}
\rm{H}^* = \arg \max_{H} \left\{ \sum_{i=1}^{H} \overline{\mathcal{T}}_{i} \leq \mathcal{T}_{QoS}\right\},
\end{equation}
where
\begin{equation}
\overline{\mathcal{T}_{i}} = (1-\flat) \cdot \overline{\mathcal{W}}_{h,i}  + \flat \cdot  \overline{\mathcal{W}}_{\ell,i} + \overline\psi_i \,+\,\tau_i ,
\end{equation} 
 $ \overline{\mathcal{W}}_{h,i} $ and $ \overline{\mathcal{W}}_{\ell,i} $ are respectively given by \eqref{WH1stM_1} and \eqref{WL1Eq} for each $i=1,2,\cdots H$, $\psi_{i}$ is the service time random variable at switch $i$, and  $\tau_i$  is the propagation delay between switch $i-1$ and switch $i$. In the special case where all switches are identical, we obtain the simpler expression:
\begin{equation}
\rm{H}^* \approx \frac{\mathcal{T}_{QoS}}{\overline{\mathcal{T}}},
\end{equation}
with $\overline{\mathcal{T}} = \overline{\mathcal{T}}_i,~\! i=1,2,\cdots K$. To draw qualitative insights, we note in the latter case that if traffic is limited, i.e. $\rho\approx 0$, then $\rm{H}^*$ is approximately given by the first order Taylor expansion: 
\begin{equation}
\rm{H}^* \approx \frac{\mathcal{T}_{QoS}}{\overline\psi + \bar\tau}\;\cdot\;\Big(1-\rho\;\frac{\overline{\mathcal{R}}}{\overline\psi + \bar\tau}\Big)
\end{equation}
Here, $ \bar \psi + \bar \tau$ is the minimum average delay at a node regardless of traffic utilization. Hence, the impact of increasing traffic utilization in IoT application is largely influenced by the ratio of residual service time $\overline{\mathcal{R}} = {\overline{\mathcal{X}^2}}/{(2\overline{\mathcal{X}})}$ to such minimum average delay. 

\subsection{Blocking Probability}

To obtain the average number of packets that exceed a predefined threshold $\mathcal{T}_{\rm{QoS}}$, we approximate the sum of all service times and waiting times experienced by a packet from source to sink by a normal distribution \cite{LCLT}. We also know the first and second moments of the random variables as derived earlier. We are interested to obtain the end-to-end delay $\mathfrak{D}=\sum_{i=1}^H \mathcal{W} + \sum_{i=1}^H  \mathcal{X}_i$ for both high-priority traffic $\mathfrak{D}_h$ and low-priority traffic $\mathfrak{D}_\ell$. Thus, we have
\begin{equation}
\mathfrak{D}_p \sim \mathcal{N}(\mu_p, \sigma_p)
\end{equation}
where
\begin{equation}
\mu_p = \sum_{i=1}^H \overline{\psi}_{i} + \sum_{i=1}^H \overline{\omega}_{p, i},
\end{equation}
\begin{equation}
\sigma_p^2 = \sum_{i=1}^H \text{Var}(\psi_{i}) + \sum_{i=1}^H \text{Var}(\omega_{p, i}),
\end{equation}
where $\psi_i$ denotes service time at node $i$ and $\omega_{p,i}$ denotes waiting time at node $i$ for traffic with priority $p\in\{\ell, h\}$.

Therefore, the final desired probability of arriving later than $\mathcal{T}_{\rm{QoS}}$ is a weighted sum according to whether a data packet is scheduled in the high-priority queue or the low-priority queue:
\begin{eqnarray}
\rm{Pr}[\mathfrak{D}>\mathcal{T}_{\rm{QoS}}] &\approx & \frac{1}{2}~\!(1-\flat)\left[ 1 - \rm{erf}\!\left(\frac{1}{\sqrt{2}} \cdot \frac{\mathcal{T}_{\rm{QoS}}-\mu_h}{\sigma_h}\right)\right]\nonumber\\
&+& \frac{1}{2}~\! \flat \left[ 1 - \rm{erf}\!\left(\frac{1}{\sqrt{2}} \cdot \frac{\mathcal{T}_{\rm{QoS}}-\mu_\ell}{\sigma_\ell}\right)\right]
\end{eqnarray}
where $\text{erf}(\cdot)$ is the error function~\cite[Eq.(8.250.1)]{Gradshteyn_Ryzhik_Book}.

\section{Numerical Results}
\label{sec:res}
Emulations are conducted by using Mininet emulator \cite{mininet} which uses real virtual hosts, POX-eel controllers \cite{pox}, and real software switches, i.e., OVS switches \cite{openvswitch}. The system characteristics of the used machine are Ubuntu 14.04 LTS installed on 16 x (2.5GHz-Intel Xeon CPU E5-2680v3), and the memory is 128 GiB. Iperf is used to generate MF and EFs of sizes 100KB and 100MB\footnote{The common block sizes in Hadoop and MapReduce algorithms are 64MB and 128MB \cite{pox}.}, respectively.

Emulated DCN topology has six spine switches and twelve leaf each with has 25 hosts, i.e., 300 hosts in total. Even though we would like to use a 10 Gbps and 100 Gbps for coaxial cables and FSO/fiber links, we scaled down all the link capacities by a factor of 10 (i.e.,  1 Gbps for coaxial and 10 Gbps for FSO/fiber links) since Mininet is limited by the processing capacity of the host machine where we implement emulations. Each FSO link consists of 4 wavelengths, which are realized as virtual links in Mininet. Since the emulator is limited in DCN size, optical channel gains are not distinguishably different, thus, assumed to be identical.

\subsection{Workloads}

We use MapReduce to mimic workloads of real DCNs, which its shuffle-phase communication pattern has $k$ servers from every rack communicate with another $k$ servers in a different rack. For instance, the hosts in $R_{i}$ have been divided into five sets and every set has $k$ servers, let's say four mice and one elephant. Each $k$ is communicating with  $k$ servers of rack $j$, $j \neq i$, different than other $k$s of the same rack. 

\subsection{Routing Algorithms}

\textit{Equal-Cost-Multi-Path routing (ECMP)~\cite{ecmp}} is a widely used DCN routing method which uses the packet header information, such as the IP/MAC addresses and TCP port numbers, as a key for a hash function. Throughout this section, ECMP refers to traditional DCNs with coaxial cables. The outgoing path is the output hash value modulo the number of outgoing paths. This strategy splits the flows among available paths. Since the header information for an individual flow is the same during the session, the packets of the same flow are always forwarded via the same path; maintaining the packet orders and avoiding flow bifurcation. We used OVS \texttt{bundle} command with \texttt{symmetric\_l4} hash function in ECMP algorithm.

\textit{ECMP-FSO} is an adaptation of ECMP to WDM-FSO case where wavelengths equally share the available FSO link capacity. That is, ECMP-FSO does not have any flexibility due to the lack of an intensity allocation mechanism and thus equivalent to WDM-fiber links. In FSO, the speeds of the links are in order of magnitude faster than non-FSO links. In this evaluation, we use a factor of ten which means the FSO link is 10$\times$ faster than traditional  DCN links. In this routing method, the link capacity is equally divided between the wavelength which means the capacity of every wavelength is fixed to 2.5 Gbps. Each flow was assigned to a single wavelength. However, when flows are more than the available number of lightpaths, the packets of the waiting flows are enqueued until a lightpath is available for transmission.

\textit{TG-FSO} is the proposed algorithm where the intensity is first allocated to the R2R mice-flow-wavelengths to meet MF demands. The remaining intensity is used by EFs as explained in Section \ref{sec:allocation}.

\begin{figure*}[htbp!]
    \centering
    \begin{subfigure}[b]{0.32\textwidth}
        \includegraphics[width=\textwidth]{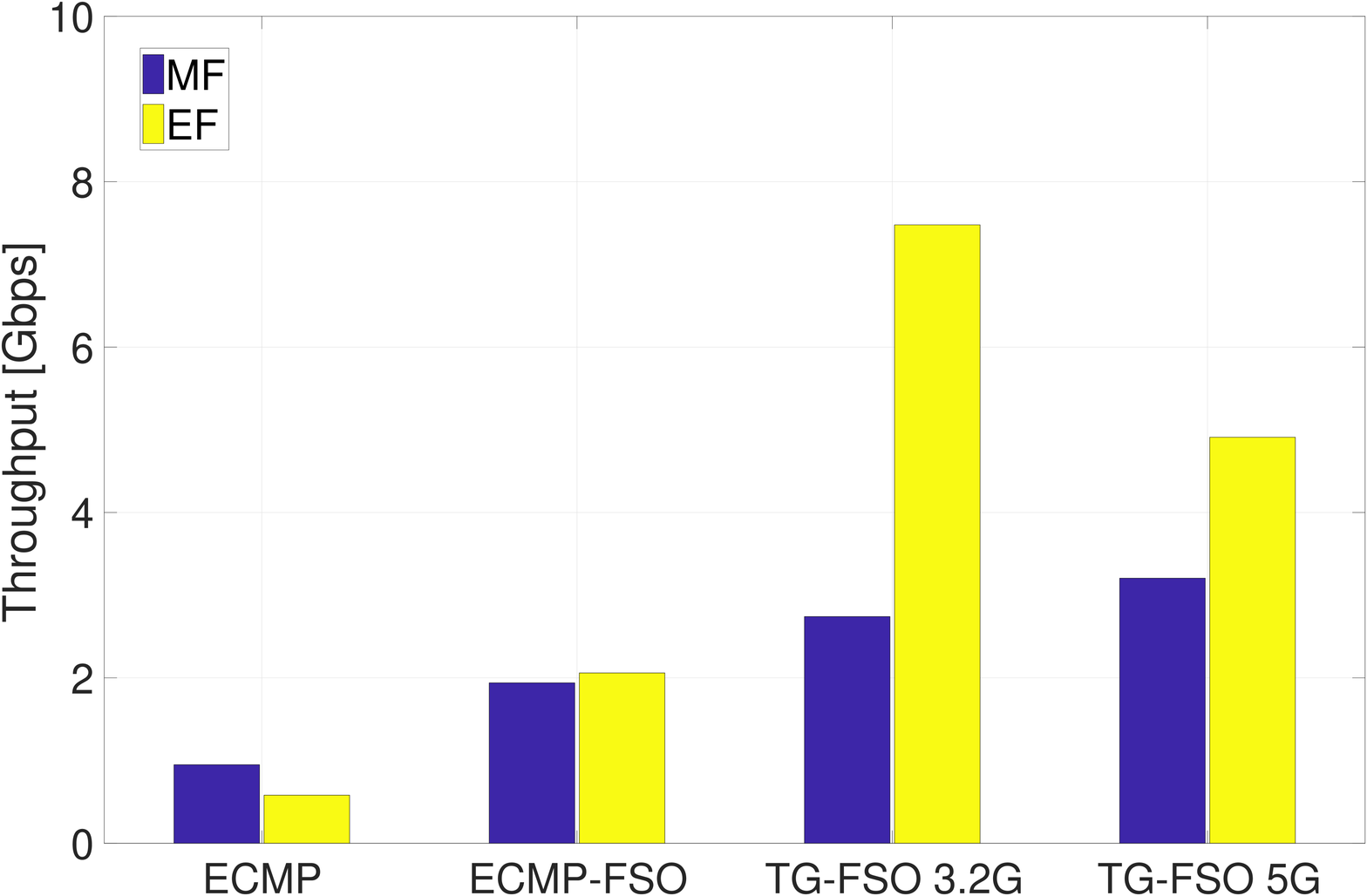}
        \caption{Throughput.}
\label{fig::throughput}
    \end{subfigure}
    \begin{subfigure}[b]{0.32\textwidth}
        \includegraphics[width=\textwidth]{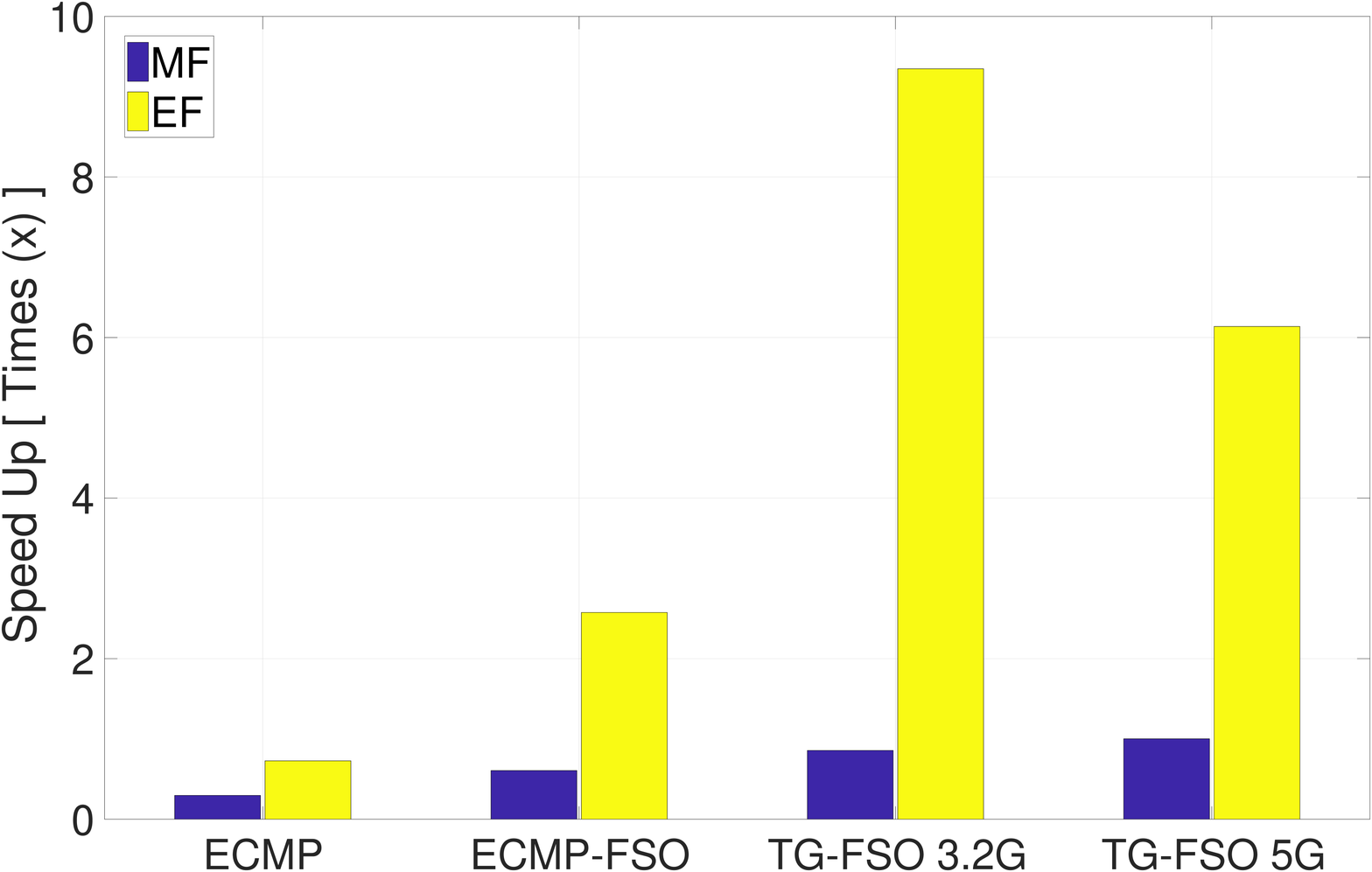}
        \caption{FCTs}
\label{fig::times}
    \end{subfigure}
    \begin{subfigure}[b]{0.32\textwidth}
        \includegraphics[width=\textwidth]{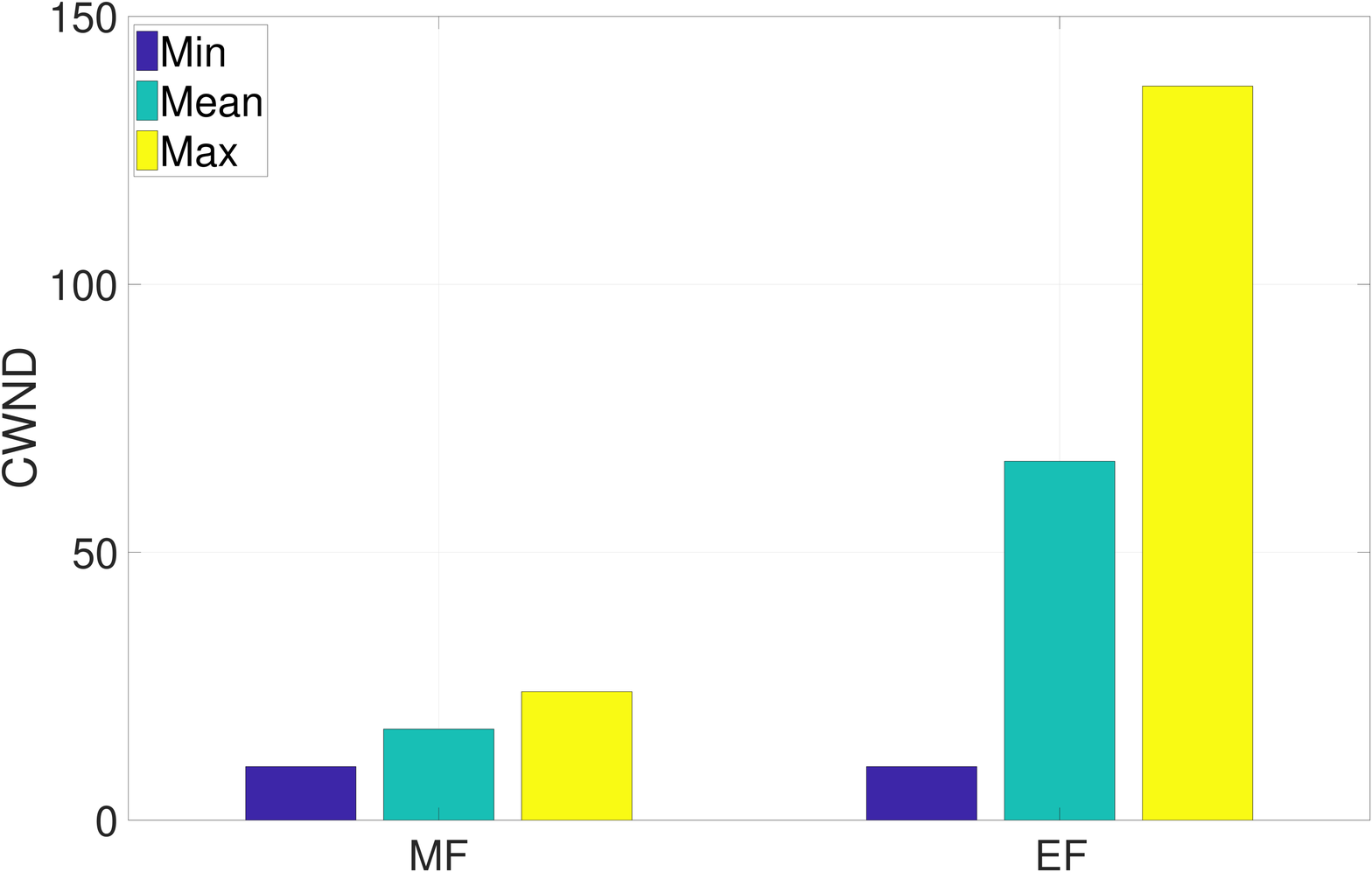}
        \caption{CWNDs}
\label{fig::cwnd}
    \end{subfigure}
    \caption{Comparison of different throughputs in terms of throughput, FCTs, and CWNDs.}\label{fig:thr_fct}
\end{figure*}

   \subsection{Network Throughput Results}
Since $k$ is set to four, we have $4 \times 100$ KB MF in every R2R communication. Hence, the needed capacity by R2R groomed MF is $3.2$ Gbps with a $\tau=1$ ms service duration.  Due to the TCP behavior; a window of packets per RTT, and the waiting time in the link buffer, some of the flow finished after $\tau$. In average 86\% of the flows complete before $\tau$. However, when the wavelength capacity increased to 5 Gbps, 100\% of the flows satisfied the time constraint. The flow completion time of EF flows achieved about $9 \times$ the needed time, i.e., one second, for the transmitted flows. The throughput and flow completion time (FCT) results are displayed in Fig.~\ref{fig::throughput} and \ref{fig::times}, respectively. 

At all configurations, proposed TG-FSO algorithm outperforms the ECMP and ECMP-FSO cases. The achievable throughput of the proposed algorithm (3.2 Gbps) is about 3.4 times the ECMP algorithm and about 1.65 times the ECMP-FSO. We should emphasize that the sustained capacity for MF is 10 Gbps and 3.2/5 Gbps for  ECMP-FSO and TG-FSO, respectively. Therefore, the transmission power consumption of  3.2/5 Gbps TG-FSO is lower than ECMP-FSO by about 3.2/2 times, respectively. For the EFs, on the other hand, the highest achievable average throughput was 7.47 Gbps which is about 13 and 3.6 times the ECMP and ECMP-FSO throughputs, respectively. However, when we increased the capacity of the allocated wavelength for MF from 3.2 Gbps to 5Gbps to satisfy the flow completion time constraint, i.e., 1 ms, the throughput of EF was decreased from 7.2 Gbps to about 5Gbps.

Due to the low capacity of the traditional data center network, the evaluated flows are completed after the time constraint on average. \textcolor{black}{Accordingly, Fig. \ref{fig::times} shows the FCT speed up of the different cases concerning the requested maximum, $\tau$.} However, the transmitted flows by using other algorithms in the same experiment complete before the time constraint. For instance, EFs in TG-FSO 3.2 Gbps complete the service nine times faster before the required service duration. We found that our algorithm in all of its versions satisfied the time constraint even if we increase the flow size or decrease the time constraint in order of magnitudes, i.e., $>$ 100MB or $\tau<1$  second. 

Even though we had to set FSO link capacity to 10 Gbps due to the Mininet restrictions,  Ciaramella et al. tested an outdoor 32 wavelength WDM-FSO system over several hundred meters and recorded 40 Gbps wavelength capacity \cite{Ciaramella2009128}. Accordingly, the potential of the proposed design can be understood better when the 13 times performance enhancement is scaled up to a higher number of wavelengths and capacities.

\subsection{Network Delay Results}
In this part of evaluations, we need to study the positive and negative impacts of the suggested queuing discipline on the performance of EF as well as mission-critical, CF, traffic. Accordingly, we selected the leaf-spine lightpaths that were dedicated to forward elephant flows and configured them with two priority queues. The high priority queue was dedicated for CFs, while the low priority was used for elephant flows. The system forwards the elephant with the full capacity as long as the high priority queue has no waiting packets.  \textcolor{black}{Also, we maintained the well-known data center traffic ratio, where 80\% of transmitted flows is MFs and 20\% is EFs \cite{bai2015information, zhang2017resilient}, i.e., $\flat=0.2$. In the traffic characteristics of social DCN, the authors of \cite{Roy2015} found that the majority of bytes during a sub-second interval are carried by large flows. Also, they found that the traffic pattern remains stable for a long period (i.e., days) and 57.5\% of the total DCN traffic is between specific racks, which is aligned with a recent study in Microsoft DC \cite{ghobadi2016projector}.}

We collected the performance figures during different link capacities, 25\%, 50\%, 75\%, and 100\%, respectively. The capacity of the lightpath is 10Gbps. We collected the average response time, the number of packets in the queue and the size of CWND of all flows transmitted through that lightpath. We used \texttt{TCP Probe}, which is a well-known tool in network measurement, to get the sender and receiver IP addresses, time of transmitting, packet length, round-trip time and CWND of all flows. Also, we logged the statistics of the low and high priority queues of the evaluated lightpath for every one millisecond. Finally, we compared the mathematical model and the collected Mininet results. 
\begin{figure}[t!]
    \centering
    \begin{subfigure}[b]{0.45\textwidth}
        \includegraphics[width=\textwidth]{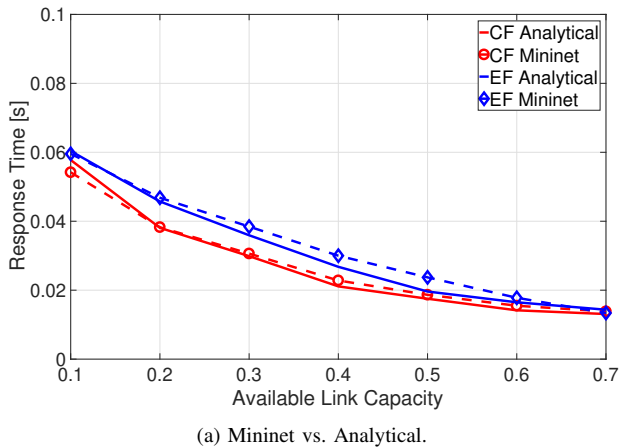}
        \caption{Mininet vs. Analytical.}
\label{fig::responsetime}
    \end{subfigure}
    \begin{subfigure}[b]{0.45\textwidth}
        \includegraphics[width=\textwidth]{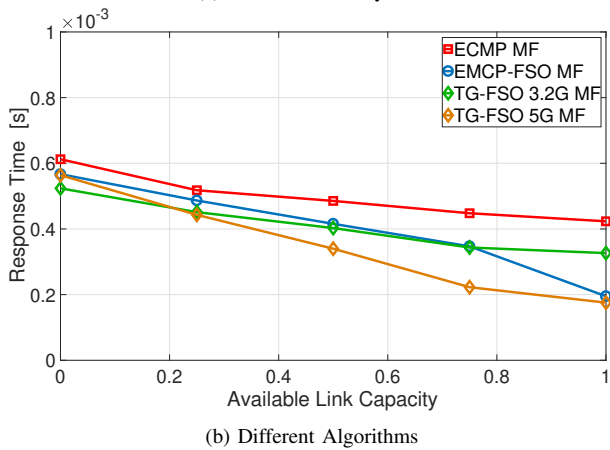}
        \caption{Different Algorithms}
\label{fig::rtmice}
    \end{subfigure}
    \caption{Response time comparison for (a) Mininet vs. analytical and (b) different algorihtms.}\label{fig:response}
\end{figure}

Before we explain our results, we need to justify our claim about the growing behavior of elephant flow CWND. To do this, we measured the mean of CWNDs of the evaluated EF and CFs. We found that the average size of EF flow CWND is 3.9$\times$ larger than the CWND of mission-critical as well as MF flows as shown in Fig.~\ref{fig::cwnd} which illustrates the minimum, mean and the maximum achieved CWNDs of EF and CFs. Fig.~\ref{fig::responsetime} shows the comparison between the response time of the mathematical model and the Mininet results. We can see from the figure how the response time increases with the increase in load. In this testing, we use the RTT, technically defined as \texttt{Short RTT}, of every transmitted flow, i.e., the CF and EF, through the evaluated lightpath. These statistics are collected from the kernel of the communicating hosts by using \texttt{TCP Probe}. The response time of MF during the evaluation with different algorithms is illustrated in Fig.~\ref{fig::rtmice}. The TG-FSO outperforms other algorithms when the load is less than 70\%. However, during high load both of the two TG-FSO settings as well as ECMP-FSO present close results.   

\begin{figure*}[t!]
    \centering
    \begin{subfigure}[b]{0.32\textwidth}
        \includegraphics[width=\textwidth]{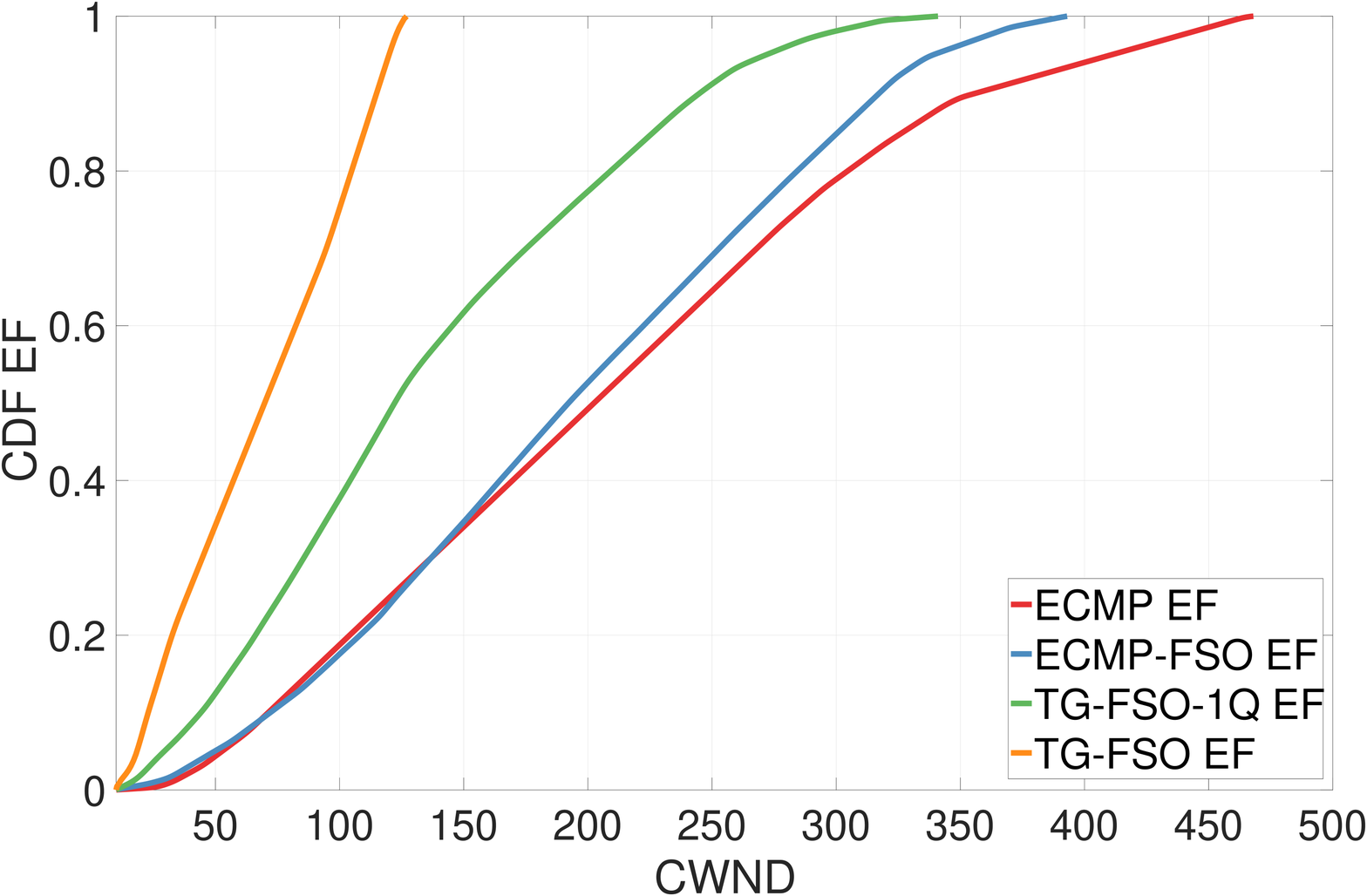}
        \caption{EFs}
\label{fig::cdfCWND}
    \end{subfigure}
    \begin{subfigure}[b]{0.32\textwidth}
        \includegraphics[width=\textwidth]{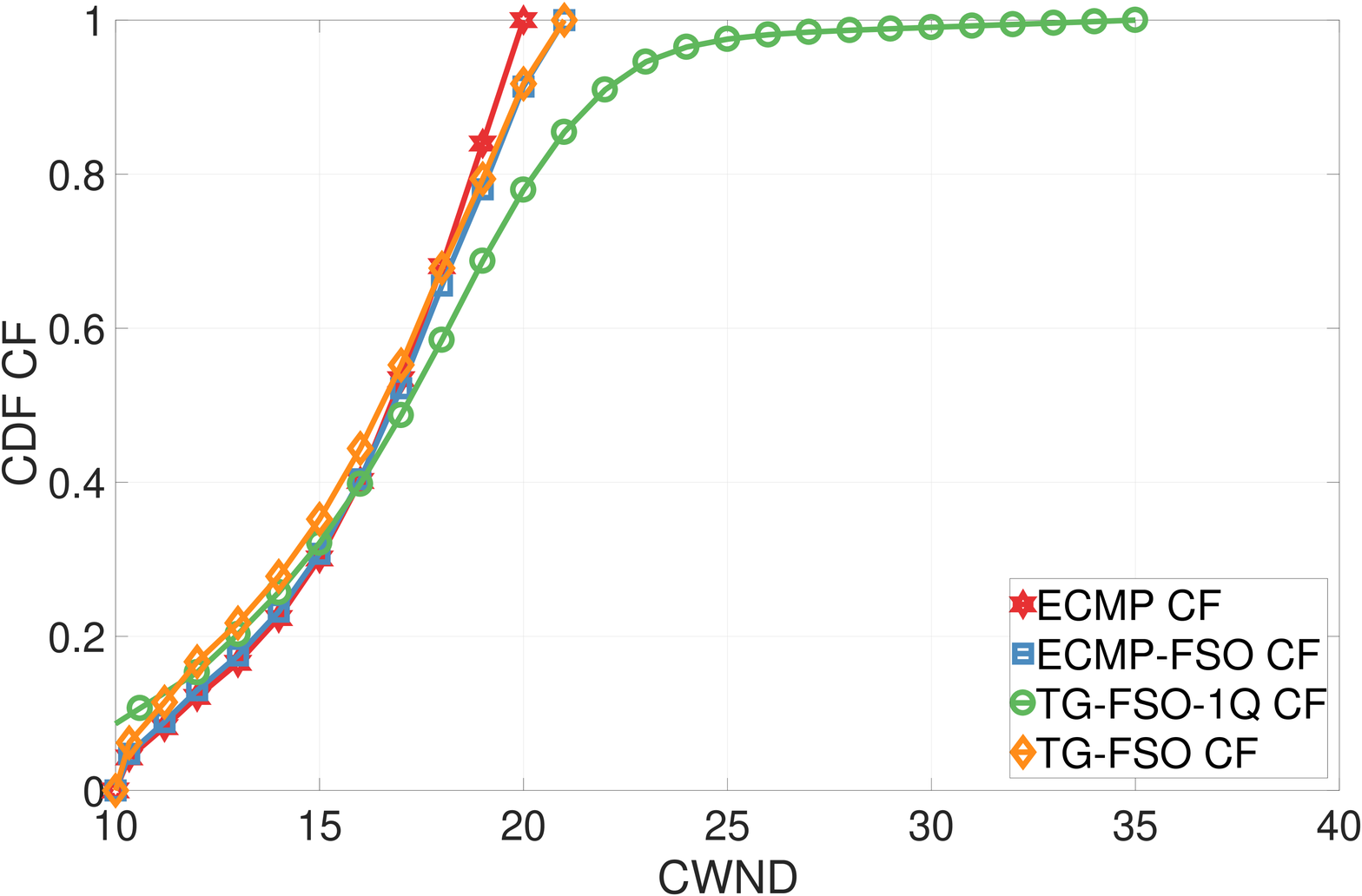}
        \caption{CFs}
\label{fig::cdfCWNDCF}
    \end{subfigure}
    \begin{subfigure}[b]{0.32\textwidth}
        \includegraphics[width=\textwidth]{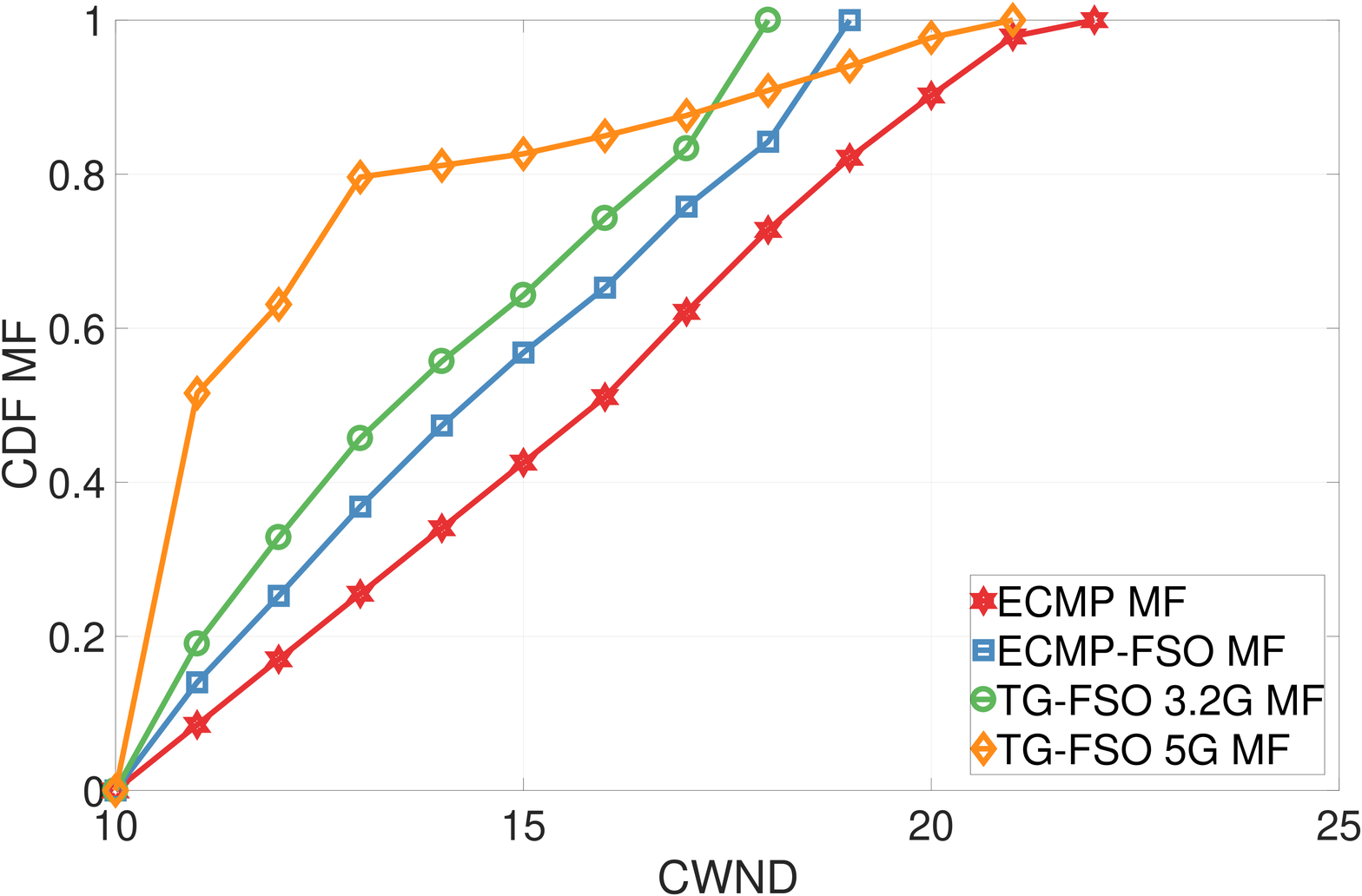}
        \caption{MFs}
\label{fig::cdfCWNDmice}
    \end{subfigure}
    \caption{CWND-CDFs for EFs, CFs, and MFs.}\label{fig:cdf}
\end{figure*}

\begin{figure*}[t!]
    \centering
        \begin{subfigure}[b]{0.32\textwidth}
        \includegraphics[width=\textwidth]{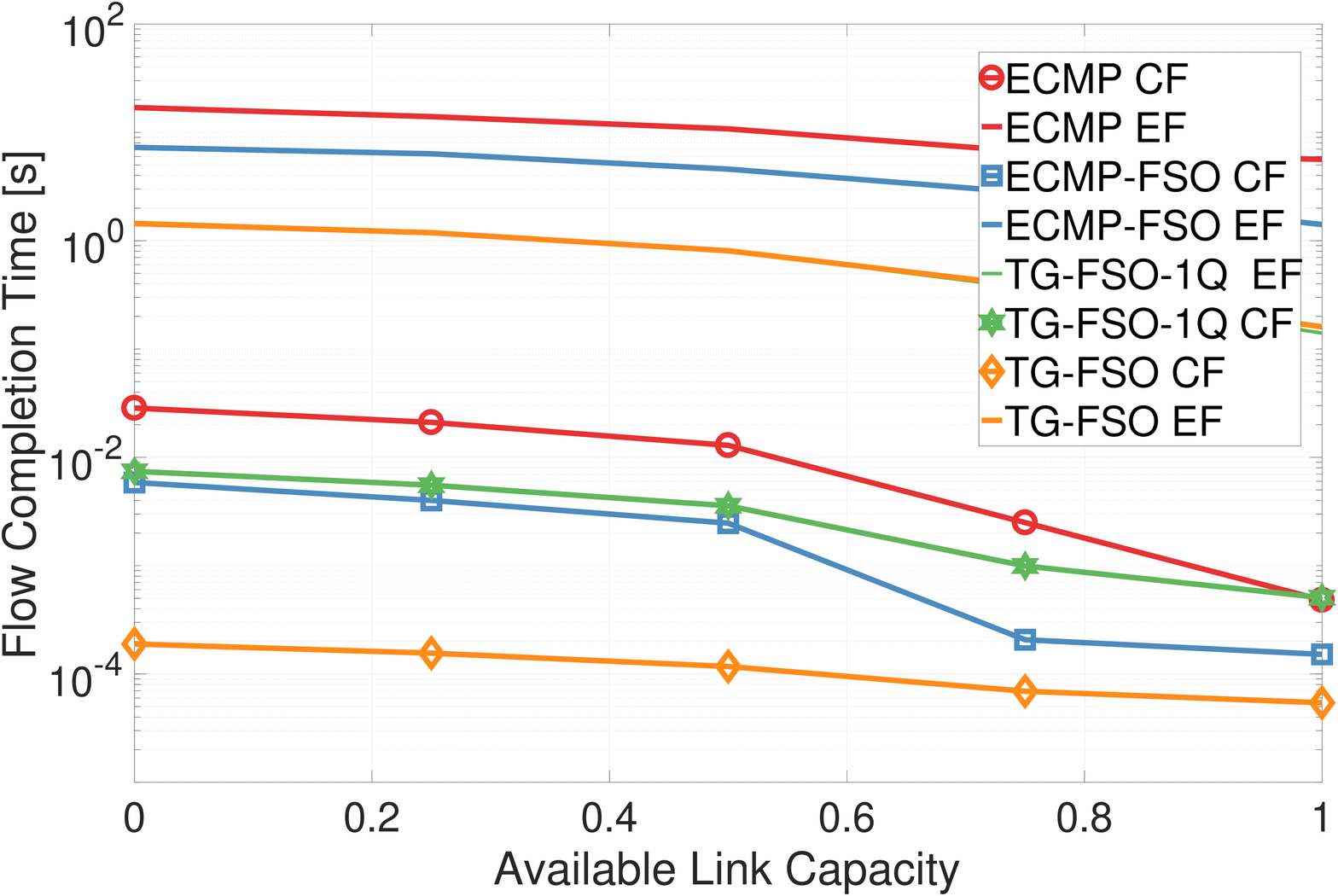}
        \caption{FCTs}
\label{fig::fct}
    \end{subfigure}
    \begin{subfigure}[b]{0.32\textwidth}
        \includegraphics[width=\textwidth]{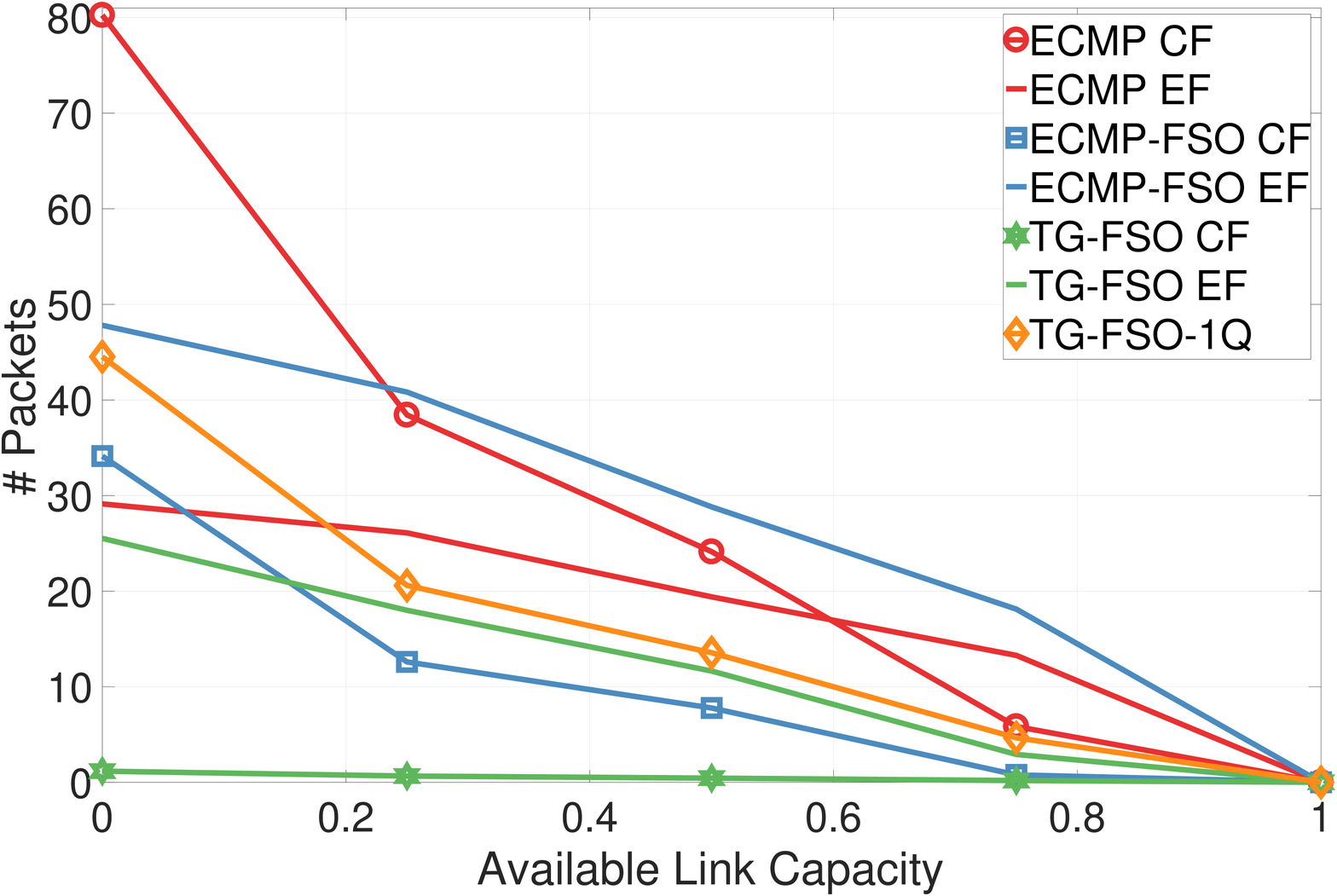}
        \caption{Number of packets.}
\label{fig::numq}
    \end{subfigure}
        \begin{subfigure}[b]{0.32\textwidth}
        \includegraphics[width=\textwidth]{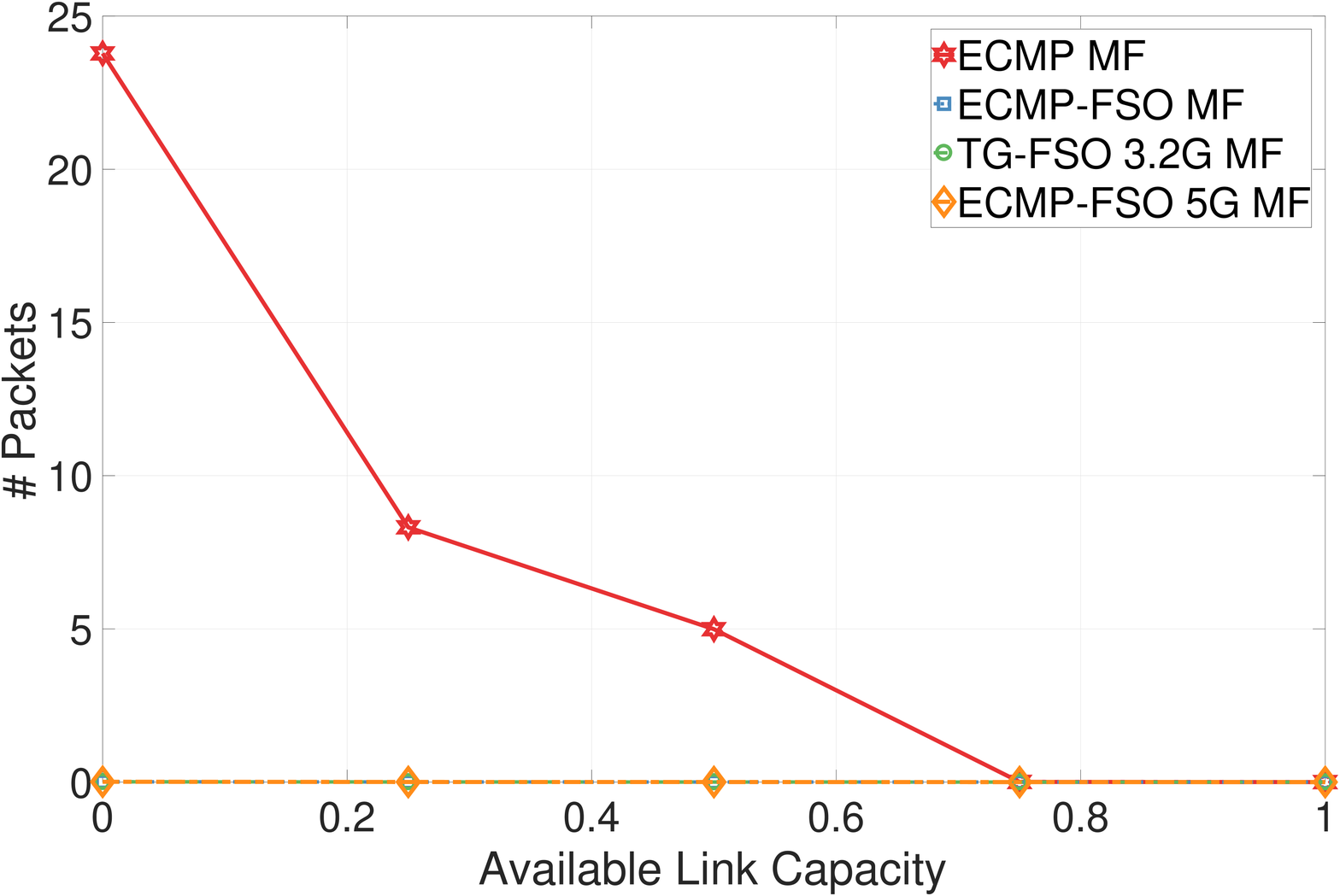}
        \caption{Queue loads}
\label{fig::numqmice}
    \end{subfigure}
    \caption{FCTs, number of packets, and queue loads for different algorithms.}\label{fig:load}
\end{figure*}

On the second part, we study the impact of the proposed queuing discipline on the CWND of EF and CFs. Fig.~\ref{fig::cdfCWND} shows the CDF of the CWND of EF and CFs. For the sake of a complete picture of evaluation, we extracted the CWND of some of MF sources; the results are displayed in Fig.~\ref{fig::cdfCWNDmice}. The CWND of mission-critical and MF flows grow at a faster rate compared to EF CWNDs. Also, the EF CWND of TG-FSO in both cases of single and double queues grew at a faster rate than other EF CWNDs. This clarifies certain phenomena; the CWND grows faster with high wavelength capacity and because the MF flows have a small amount of data to transmit they complete during the slow start phase. Although the EF has been forwarded to a low priority, its CWND grows up to about 130 packets and 470 packets in slow links, while the CWND of MF and CFs remained around its mean; 20 packets. These results manifest that the EF strives to utilize available capacity by increasing its CWND and the CFs finish before they examine the available lightpath capacity.

The impact of the suggested queuing discipline on the flow completion time of EF and CFs are illustrated in Fig.~\ref{fig::fct}. We can see how the suggested queuing discipline reduced the flow completion time of CFs by almost 2$\times$. Fortunately, the flow completion time of EF flows before and after the implementation of the queuing discipline are almost the same. Although we plot the FCT results in log scale, we didn't see a change on the FCT of EF flows. In term of waiting time, we measured the number of packets waiting in the queue. The Fig.~\ref{fig::numq} shows the number of packets waiting in the queue when the lightpaths have single and two queues. We can see in the figure that the number of packets waiting in the high priority queue is tiny, close to zero, compared to the low priority queue as well as the single queue. Since the lightpath of $TG-FSO_{1Q}$ has a single queue for all the flow classes, its the EF and CFs forwarded enqueued in the same queue. This means that the number of packets presented in the $TG-FSO_{1Q}$ curve includes both the EF and mission-critical packets. Also, the number of packets in the queue increased almost exponentially with the load. When we measured the queue occupancy in MF evaluation, we found that all the evaluated algorithms except ECMP have zero packets in the queue of the evaluated lightpath. The results are illustrated in Fig.~\ref{fig::numqmice}.

\section{Conclusions}
\label{sec:conc}
\textcolor{black}{
In this paper, we addressed the design and provisioning of mission-critical wireless DCNs from a TG perspective. To mitigate the system limitations of traditional wired DCNs, we considered a wireless approach by using hybrid optoelectronic switches and WDM capable FSO links. Contingent upon the problem formulation, we developed a fast yet high-performance sub-optimal solution which significantly improved the throughput of CFs, MFs, and EFs. Based on priority queues, the performance analyses of low and high priority flows are provided for important service characteristics including the waiting time, delay, maximum hop count, and blocking probability. By grooming the sub-wavelength traffic and adjusting the wavelength capacities according to the groomed traffic requests, numerical results clearly showed that the proposed solutions achieved significant performance enhancement by utilizing the bandwidth more efficiently, completing the flows faster than delay sensitivity requirements, and avoiding the traffic congestion by treating EFs and MFs separately.} 

% References
\bibliographystyle{IEEEtran}
\bibliography{Final.bib}

\begin{IEEEbiography}[{\includegraphics[width=1.1in,height=1.25in]{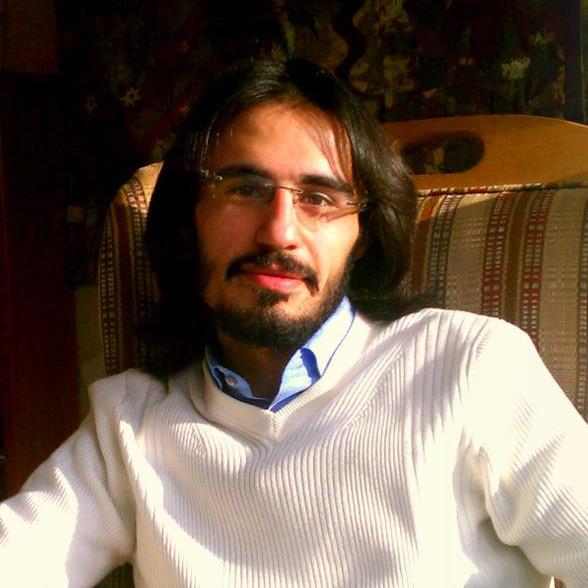}}]{Abdulkadir Celik}(S'14-M'16) received the B.S. degree in electrical-electronics engineering from Selcuk University in 2009, the M.S. degree in electrical engineering in 2013, the M.S. degree in computer engineering in 2015, and the Ph.D. degree in co-majors of electrical engineering and computer engineering in 2016, all from Iowa State University, Ames, IA. He is currently a postdoctoral research fellow at Communication Theory Laboratory of King Abdullah University of Science and Technology (KAUST). His current research interests include but not limited to 5G and beyond, wireless data centers, UAV assisted cellular and IoT networks, and underwater optical wireless communications, networking, and localization. 
\end{IEEEbiography}
\newpage

\begin{IEEEbiography}[{\includegraphics[width=1.1in,height=1.25in]{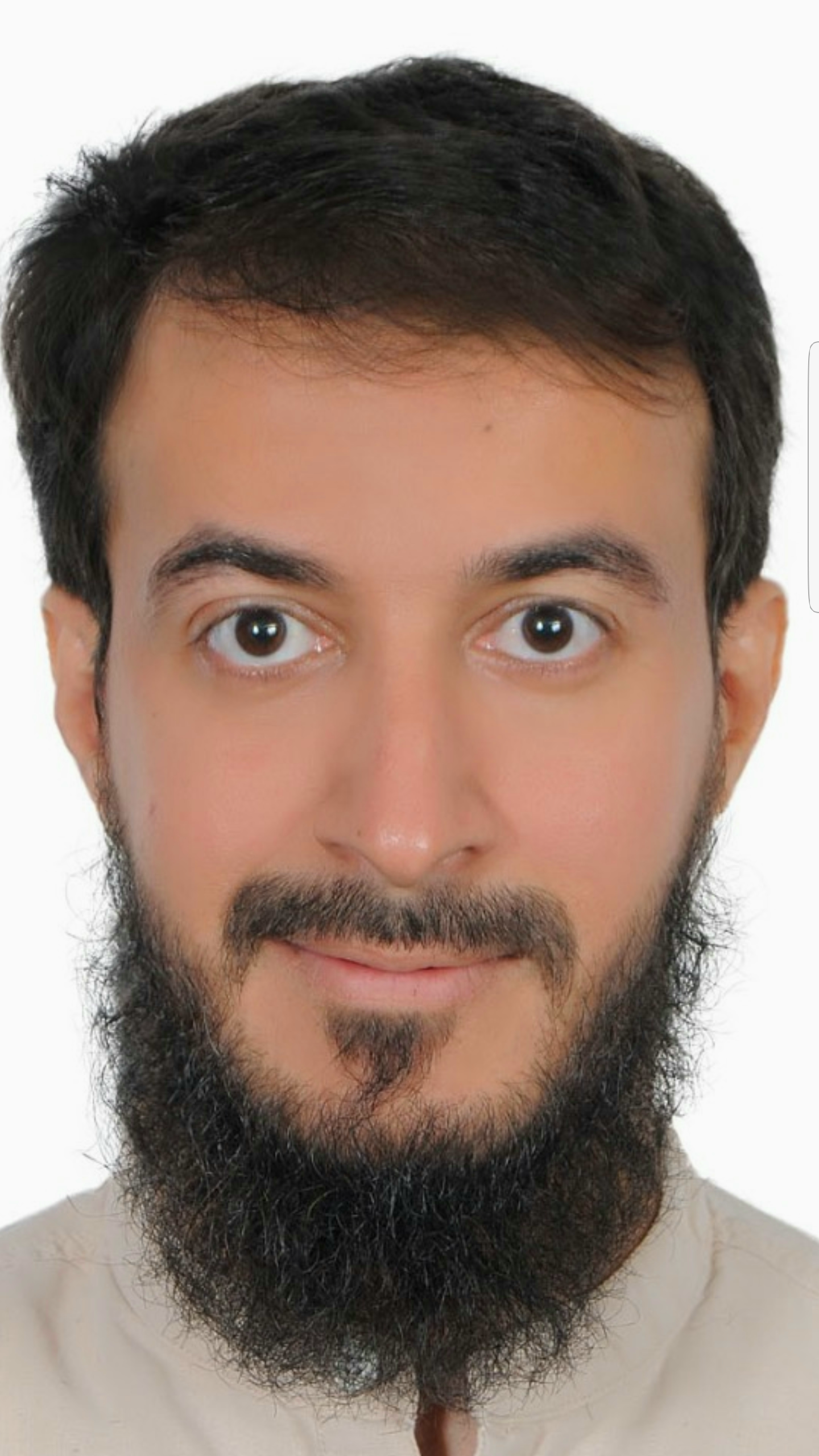}}]{Amer AlGhadhban}(SM'18)  is a Ph.D Student at Electrical Engineering department in the Computer, Electrical and Mathematical Sciences \& Engineering (CEMSE) Division at King Abdullah University of Science and Technology (KAUST). He obtained both of his degrees B.S. and M. Sc. in Computer Engineering from King Fahd University of Petroleum and Minerals (KFUPM), in (2012). He was a Cisco Academy instructor and during that he earned multiple professional certificates: Cisco Certified Network Associate (CCNA), Cisco wireless, SANS- GIAC Certified Firewall Analyst (GCFW-Gold), Certified Ethical Hacker (CEH), Security Certified Network Engineer (SCNP), and SANS local mentor. His research interests span over multiple fields in networked systems and security.

\end{IEEEbiography}

\begin{IEEEbiography}[{\includegraphics[width=1.1in,height=1.25in]{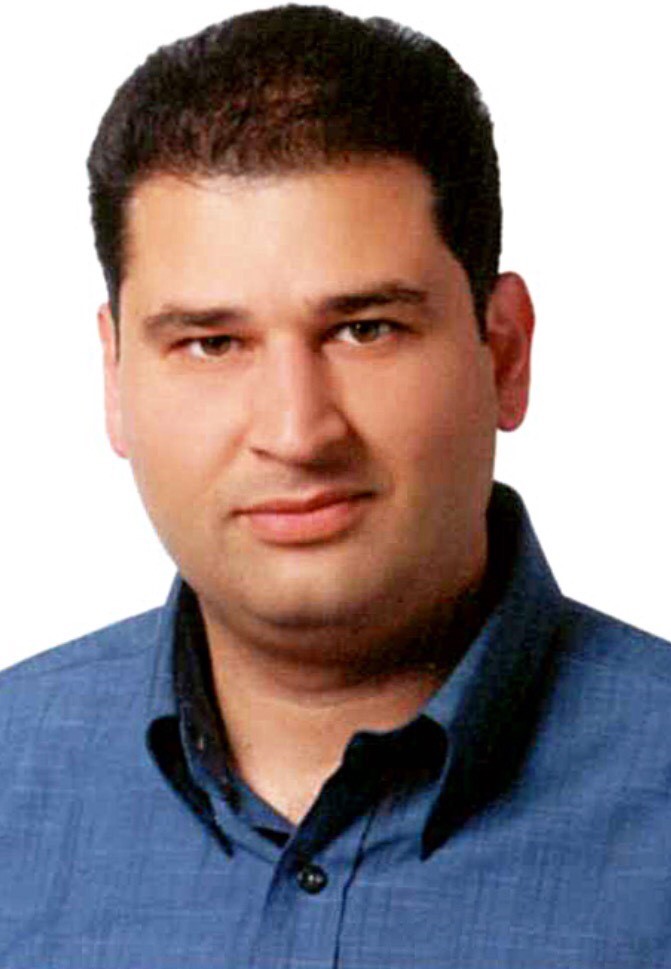}}]{Basem Shihada}(SM'12)  is an associate \& founding professor in the Computer, Electrical and Mathematical Sciences \& Engineering (CEMSE) Division at King Abdullah University of Science and Technology (KAUST). He obtained his PhD in Computer Science from University of Waterloo. In 2009, he was appointed as visiting faculty in the Department of Computer Science, Stanford University. In 2012, he was elevated to the rank of Senior Member of IEEE. His current research covers a range of topics in energy and resource allocation in wired and wireless networks, software defined networking, internet of things, data networks, network security, and cloud/fog computing. 
\end{IEEEbiography}

\begin{IEEEbiography}[{\includegraphics[width=1.1in,height=1.25in]{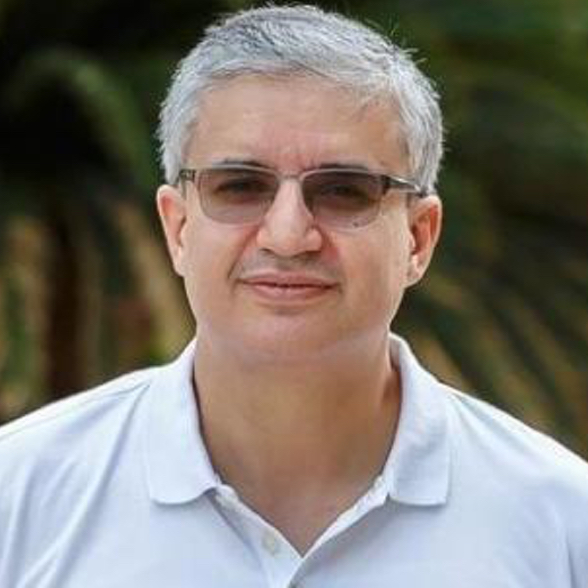}}]{Mohamed-Slim Alouini} (S'94-M'98-SM'03-F'09) was born in Tunis, Tunisia. He received the Ph.D. degree in Electrical Engineering from the California Institute of Technology (Caltech), Pasadena, CA, USA, in 1998. He served as a faculty member in the University of Minnesota, Minneapolis, MN, USA, then in the Texas A\&M University at Qatar, Education City, Doha, Qatar before joining King Abdullah University of Science and Technology (KAUST), Thuwal, Makkah Province, Saudi Arabia as a Professor of Electrical Engineering in 2009. His current research interests include the modeling, design, and performance analysis of wireless communication systems.
\end{IEEEbiography}

\end{document}